\documentclass[10pt]{article}
\usepackage{amsmath}
\usepackage{amssymb}
\usepackage{eufrak}
\usepackage{oldgerm}
\usepackage{graphicx}
\usepackage{psfrag}
\usepackage{cite}
\begin{document}
%%%%%%%%%%%%%%%%%%%%%%%%%%%%%%%%%%%%%%%%%%%%%%%%%%%%%%%%%%%%%%%%%%%%%%%%
%%%%%%%%%%%%%%%%%%%%%%%%%%%%%%%%%%%%%%%%%%%%%%%%%%%%%%%%%%%%%%%%%%%%%%%%
\title{Remodeling of biological tissue:
Mechanically induced reorientation of a transversely isotropic
chain network}
%%%%%%%%%%%%%%%%%%%%%%%%%%%%%%%%%%%%%%%%%%%%%%%%%%%%%%%%%%%%%%%%%%%%%%%%
\author{Ellen Kuhl\\Chair for Applied Mechanics\\
         University of Kaiserslautern,
         D-67653 Kaiserslautern, Germany\\
{\tt ekuhl@rhrk.uni-kl.de}\\[20 pt]
Krishna Garikipati \\Department of Mechanical Engineering,\\
         University of Michigan,
         Ann Arbor, MI 48109, USA\\
         {\tt krishna@engin.umich.edu}\\[20 pt]
Ellen M. Arruda\\Department of Mechanical
Engineering\\Macromolecular Science and Engineering
Program\\University of Michigan,
         Ann Arbor, MI 48109, USA\\{\tt arruda@umich.edu}\\[20 pt]
Karl Grosh\\Department of Mechanical Engineering,\\
         University of Michigan,
         Ann Arbor, MI 48109, USA\\{\tt grosh@engin.umich.edu}}
\maketitle

\begin{abstract}
A new class of micromechanically motivated chain network models for soft biological
tissues is presented. On the microlevel, it is based on the statistics of
long chain molecules.
A wormlike chain model is applied to capture the
behavior of the collagen microfibrils. On the macrolevel, the network of
collagen chains is represented by a transversely isotropic eight chain unit cell
introducing one characteristic material axis.
Biomechanically induced remodeling is captured by
allowing for a continuous reorientation of the predominant unit cell axis driven
by a biomechanical stimulus. To this end, we adopt the gradual alignment of
the unit cell axis with the direction of maximum principal strain.
The evolution of the unit cell axis' orientation is governed by a first-order
rate equation. For the
temporal discretization of the remodeling rate equation, we suggest an
exponential update scheme of Euler-Rodrigues type. For the spatial
discretization, a finite element strategy is applied which introduces the
current individual cell orientation as an internal variable on the integration
point level. Selected model problems are analyzed to illustrate the basic
features of the new model. Finally, the presented approach is applied to the
biomechanically relevant boundary value problem of an in vitro engineered
functional tendon construct.
\end{abstract}
%%%%%%%%%%%%%%%%%%%%%%%%%%%%%%%%%%%%%%%%%%%%%%%%%%%%%%%%%%%%%%%%%%%%%%%%
%%%%%%%%%%%%%%%%%%%%%%%%%%%%%%%%%%%%%%%%%%%%%%%%%%%%%%%%%%%%%%%%%%%%%%%%

%%%%%%%%%%%%%%%%%%%%%%%%%%%%%%%%%%%%%%%%%%%%%%%%%%%%%%%%%%%%%%%%%%%%%%%%
\newtheorem{remark}{Remark}
\newcommand{\beq}{\begin{equation}}
\newcommand{\eeq}{\end{equation}}
\newcommand{\bea}{\begin{eqnarray}}
\newcommand{\eea}{\end{eqnarray}}
\newcommand{\D}  {\displaystyle}
\newcommand{\Divg}{\mbox{\rm{Div}}\,}
\def\sca   #1{\mbox{#1}{}}
\def\mat   #1{\mbox{\bf #1}{}}
\def\vec   #1{\mbox{\boldmath $#1$}{}}
\def\ten   #1{\mbox{\boldmath $#1$}{}}
\def\scas  #1{\mbox{{\scriptsize{${\rm{#1}}$}}}{}}
\def\vecs  #1{\mbox{{\boldmath{\scriptsize{$#1$}}}}{}}
\def\tens  #1{\mbox{{\boldmath{\scriptsize{$#1$}}}}{}}
%%%%%%%%%%%%%%%%%%%%%%%%%%%%%%%%%%%%%%%%%%%%%%%%%%%%%%%%%%%%%%%%%%%%%%%%
%%%%%%%%%%%%%%%%%%%%%%%%%%%%%%%%%%%%%%%%%%%%%%%%%%%%%%%%%%%%%%%%%%%%%%%%
\section{Introduction}
%%%%%%%%%%%%%%%%%%%%%%%%%%%%%%%%%%%%%%%%%%%%%%%%%%%%%%%%%%%%%%%%%%%%%%%%
%%%%%%%%%%%%%%%%%%%%%%%%%%%%%%%%%%%%%%%%%%%%%%%%%%%%%%%%%%%%%%%%%%%%%%%%
% motivation
%%%%%%%%%%%%%%%%%%%%%%%%%%%%%%%%%%%%%%%%%%%%%%%%%%%%%%%%%%%%%%%%%%%%%%%%
Collagen is a fibrous protein secreted by connective tissue
cells. The distinguishing feature of a typical collagen molecule is its long,
stiff, triple-stranded helical structure, in which three collagen polypeptide
chains, the so-called $\alpha$ chains, are wound around one another in a
rope-like superhelix. So far, about 25 distinct collagen $\alpha$ chains have been
identified. The $\alpha$ chains that make up type \nolinebreak[4]I
collagen are by far the most common. Type I collagen is a fibril-forming collagen
which is present in nearly all connective tissues such as
bone, skin, tendons or ligaments. After being secreted
into the extracellular space, collagen molecules assemble into
collagen microfibrils. These are sparsely cross-linked
higher order polymers which are about 10-300 nm in diameter and several
hundreds of micrometers long, see e.g. the fundamental textbook on cell biology by
Alberts et al. \cite{alberts02} or the rather mechanically oriented textbooks on
soft biological tissues by
Fung \cite{fung93},
Cowin \& Humphrey \cite{cowin01},
Holzapfel \& Ogden \cite{holzapfel03},
Humphrey \cite{humphrey02a} and
Humphrey \& Delange \cite{humphrey04}.\\
%%%%%%%%%%%%%%%%%%%%%%%%%%%%%%%%%%%%%%%%%%%%%%%%%%%%%%%%%%%%%%%%%%%%%%%%
% statistics of long chain molecules -> rubber
%%%%%%%%%%%%%%%%%%%%%%%%%%%%%%%%%%%%%%%%%%%%%%%%%%%%%%%%%%%%%%%%%%%%%%%%
Polymer chains have many conformations of nearly equal energy.
Perturbing the chains away from their equilibrium conformations typically
generates entropic forces that oppose these perturbations.
This is the basis for entropy based elasticity.
Since the number of different
configurations which a long chain molecule may assume
is very large, the treatment of each of them individually would be
a complex, maybe even unmanagable, task. Long chain molecules are thus commonly described by
statistical mechanics, a concept which was 
originally developed in the context of entropic rubber elasticity by
Kuhn \cite{kuhn34,kuhn36} or
Kuhn \& Gr\"un \cite{kuhn42},
see also the textbooks by
Flory \cite{flory69} or Treloar \cite{treloar75}.
As pointed out by
Boyce \cite{boyce96} and
Boyce \& Arruda \cite{boyce00},
statis\nolinebreak[4]tical, microscopically motivated models are typically characterized
by a limited number of
well-defined, physically-motivated material parameters.
This is in contrast to the
macroscopic phenomenological models documented e.g. by
Ogden \cite{ogden72,ogden84},
Treloar \cite{treloar75,treloar76} or
Holzapfel \cite{holzapfel00}.\\
Collagen chains are extremely rich in proline and glycine. While the former
stabilizes the helical conformation of each $\alpha$ chain, the latter
allows the three $\alpha$ chains to pack tightly together in the final
collagen superhelix. Unlike polymer chains in rubber, which are of rather
{\it uncorrelated} nature, collagen chains in biological tissues thus have to be
classified as {\it correlated} chains from a statistical point of view.
Rubber chains are typically characterized as freely jointed chains,
the configuration of which resembles a random walk,
whereas biological chains
correspond to ``wormlike chains'' with a smoothly varying curvature, see
Kratky \& Porod \cite{kratky49}. Traditionally,
the wormlike chain model has been applied to describe the behavior of the
DNA double helix, e.g.
Marko \& Siggia \cite{marko95} or
Bustamante et al. \cite{bustamante03}. Only recently, the wormlike chain approach has
been adopted to simulate the constitutive behavior of the collagen triple helix
by Bischoff et al. \cite{bischoff02,bischoff02a} and
Garikipati et al. \cite{garikipati04}.\\
%%%%%%%%%%%%%%%%%%%%%%%%%%%%%%%%%%%%%%%%%%%%%%%%%%%%%%%%%%%%%%%%%%%%%%%%
% chain network models 
%%%%%%%%%%%%%%%%%%%%%%%%%%%%%%%%%%%%%%%%%%%%%%%%%%%%%%%%%%%%%%%%%%%%%%%%
From the standpoint of polymer structures, rubber as well as soft biological
tissue consists of a complex three-dimensional network of polymer
chains laterally attached to one another at occasional points along their
lengths. To account not only for the behavior of the individual chains as such,
but also for the characteristic cross-link effects of the network structure,
a number of different chain network models have been proposed over the past 60
years. The common feature of all these network models is a
characteristic unit cell which is assumed to provide an adequate representation
of the underlying macromolecular network structure.
The existing chain network models can basically be classified in two
categories: {\it affine} and {\it non-affine} models.\\
The first affine network model was the three chain model by 
James \& Guth \cite{james43} and
Wang \& Guth \cite{wang52}. Relating the overall strain state in an affine
manner to the stretches of three mutually orthogonal chains, the model
generally overes\nolinebreak[4]timates the overall stiffness
when assuming that one out of these three chains is
aligned with the direction of maximum principal strain.
Moreover, due to the orthogonal arrangement of the three chains, 
cross-linking network effects are basically neglected.
To remedy these deficiencies, an affine full network model
was suggested by
Treloar \& Riding \cite{treloar79}, compare also
Wu \& van der Giessen \cite{wu93} or
Miehe \cite{miehe04}.
As its chains are distributed with equal probability over the solid angle,
the affine full network model is considerably weaker than the three chain
model since only a limited number of chains are strained up to the
locking stretch upon uniaxial deformation.\\
The first representative of the class of non-affine models was the
four chain tetrahedron model by
Flory \& Rehner \cite{flory43} and
Treloar \cite{treloar46}. It is not surprising though, that due to the
tetrahedral shape of its unit cell, the four chain model reveals
a non-physical anisotropic response.
The non-affine eight chain model based on a cubic unit cell
is maybe the most prominent isotropic chain network model today,
compare Arruda \& Boyce \cite{arruda93} or Boyce \cite{boyce96}.
Although counterintuitive at first glance, non-affine chain network models
seem to be superior over affine models in
predicting the overall response under various load cases
as they assume an instantaneous orientation of the unit cell with respect to the
principal stretch space.
Critical comparisons of the different isotropic
chain network models can be found in the classical textbook by
Treloar \cite{treloar75} or in the monographs by
Flory \cite{flory76},
Boyce \& Arruda \cite{boyce00} or
Miehe et al. \cite{miehe04}.\\
%%%%%%%%%%%%%%%%%%%%%%%%%%%%%%%%%%%%%%%%%%%%%%%%%%%%%%%%%%%%%%%%%%%%%%%%
% anisotropy / experiments in biomechanics
%%%%%%%%%%%%%%%%%%%%%%%%%%%%%%%%%%%%%%%%%%%%%%%%%%%%%%%%%%%%%%%%%%%%%%%%
It was already pointed out by Kuhn \& Gr\"un \cite{kuhn42} that,
for rubberlike materials, the assumption of isotropy is typically only
met in the small strain regime. Upon further loading, the polymer
chains were found to reorient themselves with respect to the loading
direction. There is experimental evidence that this effect of anisotropy is
even more pronounced in soft biological tissues, see e.g.
the early experiments by
Lanir \& Fung \cite{lanir74}, or the textbooks by
Fung \cite{fung93},
Holzapfel \& Ogden \cite{holzapfel03} and
Humphrey \cite{humphrey02a}.
It is widely accepted that this effect of anisotropy can be attributed
to the collageneous network structure in which the polymer chains
are typically oriented with respect to the predominant loading
direction. A classical example is provided by
Langer's lines in skin, pointing in the direction
of maximum tensile strain. Similar effects can obviously be observed
in muscles, tendons and ligaments.
A first attempt at modeling the anisotropic response of soft tissues
based on an anisotropic eight chain unit cell
has been presented recently by
Bischoff et al. \cite{bischoff00,bischoff02,bischoff02a} and
Garikipati et al. \cite{garikipati04}. In the present work, we shall
adopt the above concepts to derive a transversely isotropic chain
network model, see also Kuhl et al. \cite{kuhl04c}.
Following the ideas of the classical eight chain model,
the isotropic in-plane response is represented in a {\it non-affine}
manner, while the out-of-plane stretch is related to the overall
macroscopic strain through an {\it affine} transformation. \\
%%%%%%%%%%%%%%%%%%%%%%%%%%%%%%%%%%%%%%%%%%%%%%%%%%%%%%%%%%%%%%%%%%%%%%%%
% anisotropic growth 
%%%%%%%%%%%%%%%%%%%%%%%%%%%%%%%%%%%%%%%%%%%%%%%%%%%%%%%%%%%%%%%%%%%%%%%%
Although characteristic for fully developed tissue, 
the pronounced orientation of collagen chains is typically not present
in the embryonic state, see e.g. Calve et al. \cite{calve04}.
It is only upon mechanical loading, that the tissue exhibits a
directional strengthening due the local rearrangement of the collagen
fibers, an effect with is referred to as ``functional adaptation''
or ``remodeling'' in the
biomechanical literature, see e.g. Taber \cite{taber95}.
Again, two approaches can be distinguished, macroscopic phenomenological
concepts and micromechanically motivated strategies. The former are
typically based on the introduction of a ficticious
growth configuration and the characterization of its evolution with
respect to the undeformed configuration, see e.g.
Rodriguez et al. \cite{rodriguez94},
Lubarda \& Hoger \cite{lubarda02},
Epstein \& Maugin \cite{epstein00} and
Garikipati et al. \cite{garikipati03}.
Micromechanically motivated remodeling theories, on the contrary,
are based on the rigorous reorientation of collagen fibers, see e.g.
Cowin \cite{cowin95},
Vianello \cite{vianello96},
Sgarra \& Vianello \cite{sgarra97},
Driessen et al. \cite{driessen03}.
The present work is particularly related to the recent contribution by
Menzel \cite{menzel04}, who suggests a reorientation of the material's
principal axis with respect to the predominant loading direction.\\
Within this contribution, we suggest a {\it gradual} alignment of the
axis of transverse isotropy with respect to the direction of maximum
principal strain. 
Recall, that upon loading, the non-affine eight chain model
by Arruda \& Boyce \cite{arruda93} tacitly assumes an
{\it instantaneous} reorientation of the representative unit cell, such
that its axes always remain aligned with the principal
strain axes. Herein, the approach of {\it instantaneous} alignment
is adopted for the isotropic {\it non-affine} in-plane response,
while for the {\it affine} out-of-plane response,
we postulate a {\it gradual} reorientation.\\
%%%%%%%%%%%%%%%%%%%%%%%%%%%%%%%%%%%%%%%%%%%%%%%%%%%%%%%%%%%%%%%%%%%%%%%%
% structure of the paper
%%%%%%%%%%%%%%%%%%%%%%%%%%%%%%%%%%%%%%%%%%%%%%%%%%%%%%%%%%%%%%%%%%%%%%%%
The present manuscript is organized as follows.
Section \ref{chain} introduces the basic ideas of the micromechanics
of long chain molecules to simulate the individual collagen fibrils.
In particular, two different types of model chains
are discussed and compared: the freely jointed chain and the wormlike chain.
Section \ref{homog} then deals with the mechanics of the chain network. Based on
the concept of a representative eight chain unit cell,
we consider a transversely isotropic
network model of wormlike chain type. We then address
the issue of reorientation. Accordingly, section \ref{remod} focuses on the
continuum model of remodeling and its algorithmic realization.
The features of the single chain model, the chain network model and the reorientation model
are illustrated individually in terms of a homogeneous model problem under
uniaxial tension in each section. Finally, the overall model is used to predict
the fiber reorientation of randomly oriented collagen fibers in a
cylindrical model tendon. The suggested remodeling approach
is discussed in section \nolinebreak[4]\ref{concl}.
%%%%%%%%%%%%%%%%%%%%%%%%%%%%%%%%%%%%%%%%%%%%%%%%%%%%%%%%%%%%%%%%%%%%%%%%
\begin{remark}[Notion of affinity]
%%%%%%%%%%%%%%%%%%%%%%%%%%%%%%%%%%%%%%%%%%%%%%%%%%%%%%%%%%%%%%%%%%%%%%%%
Throughout this paper, we shall apply the notion of ``{\it affinity}'' in
the context of ``{\it affine motion}''. According to
Holzapfel \cite{holzapfel00}, affinity in this sense implies that 
changes in the length and orientation of lines marked on chains in a
network are identical to changes in lines marked on the corresponding
dimensions of the macroscopic sample. In the principal strain space,
the eight chain model can thus be classified as an affine model. For arbitrary
load cases, however, the eight chain model does not assume affine
deformation of all chains, see Boyce \& Arrunda \cite{boyce00} or
Miehe et al. \cite{miehe04}. Note that we do not use the notion
of affinity in the context of ``{\it affine junction motion}''
which it was related to earlier in the context of constrained junction
theories or phantom models, i.e. models which nicely caputre the network
behavior at moderate strains, see e.g. Flory \& Erman \cite{flory82}.
%%%%%%%%%%%%%%%%%%%%%%%%%%%%%%%%%%%%%%%%%%%%%%%%%%%%%%%%%%%%%%%%%%%%%%%%
\end{remark}

%%%%%%%%%%%%%%%%%%%%%%%%%%%%%%%%%%%%%%%%%%%%%%%%%%%%%%%%%%%%%%%%%%%%%%%%
\section{Micromechanics of a single collagen chain}\label{chain}
%%%%%%%%%%%%%%%%%%%%%%%%%%%%%%%%%%%%%%%%%%%%%%%%%%%%%%%%%%%%%%%%%%%%%%%%
We shall begin by introducing the kinematics of a single polymer model chain.
In the simplest case, this chain can be characterized through
$N$ rigid bonds of equal length $l$, the so-called Kuhn length,
see e.g. Kuhn \cite{kuhn34,kuhn36}.
Accordingly, the total stretched-out length of a chain, the contour
length, is given as $L=N\,l$.
%%%%%%%%%%%%%%%%%%%%%%%%%%%%%%%%%%%%%%%%%%%%%%%%%%%%%%%%%%%%%%%%%%%%%%%
\begin{figure}[h]
\unitlength1.0mm
\begin{picture}(139.0,52.0)
\put(0.0,
0.0){\includegraphics[width=139mm,angle=0]{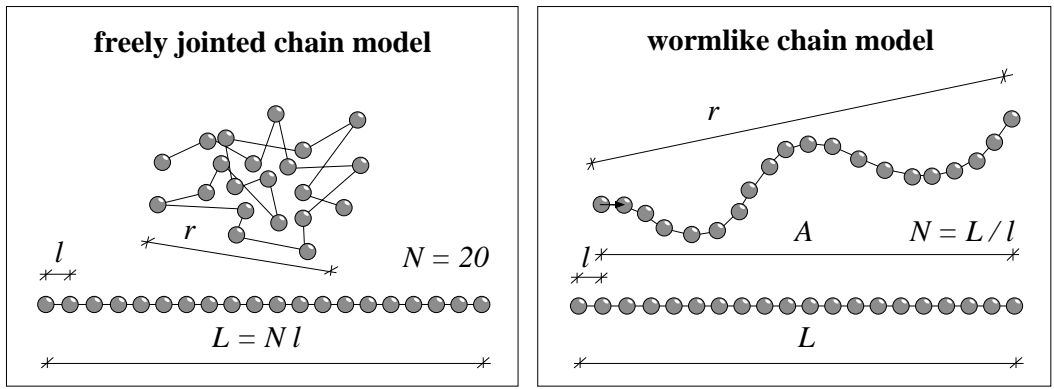}}
\end{picture}
\caption{Kinematics of individual chain models}
\label{chains_kinem}
\end{figure}
%%%%%%%%%%%%%%%%%%%%%%%%%%%%%%%%%%%%%%%%%%%%%%%%%%%%%%%%%%%%%%%%%%%%%%%%
The deformation of this model chain is then typically defined in terms of the
end-to-end distance $r$, i.e. the length of the
vector pointing from one end of the chain to the other, whereby
$0 \le r \le L$.
Alternatively, the deformation of the chain can be characterized through the
relative chain stretch $\lambda=r/L$,
i.e. the dimensionless ratio between the end-to-end length and
the contour length.
While $l$, $N$ and $L$ are constant for a particular
chain, the end-to-end length $r$ and the chain stretch
$\lambda=r/L$
change as the chain is subjected to applied forces.
Two different types of chain models can be distinguished from a
statistical point of view: uncorrelated and correlated chains,
as illustrated in figure \ref{chains_kinem}.
In what follows, we will discuss different representatives of each of these
classes, i.e. the classical single parameter
freely jointed chain and the two parameter wormlike chain.
%%%%%%%%%%%%%%%%%%%%%%%%%%%%%%%%%%%%%%%%%%%%%%%%%%%%%%%%%%%%%%%%%%%%%%%%
\subsection{Uncorrelated chains: The freely jointed chain model}
\label{micro_fjc}
%%%%%%%%%%%%%%%%%%%%%%%%%%%%%%%%%%%%%%%%%%%%%%%%%%%%%%%%%%%%%%%%%%%%%%%%
The most common model for chains is the random flight or rather freely jointed
chain model. The freely jointed chain consists of $N$ bonds of fixed bond
length $l$, whereby the directions of neighboring bonds are completely uncorrelated
in the sense that all directions for a given bond are of equal probability,
irrespective of the directions of the neighboring segments. The model is
thus characterized through one single parameter, the contour length $L=N\,l$.
Figure \ref{chains_kinem}, left,
illustrates a freely jointed chain with $N=20$ bonds.
Let us introduce the probability density $p (\lambda)$,
i.e. the probability that a chain of the contour length $L$
takes a configuration characterized through
the end-to-end length $r$. According to the classical
Boltzmann equation $s^{\scas{fjc}}=k\,\ln \, (p)$, the entropy
$s^{\scas{fjc}}$ of a single chain can be expressed in terms of the
Boltzmann constant $k=1.38 10^{-23}$J/K
and the probability density $p$. For purely entropic chains,
the free energy $\psi^{\scas{fjc}}$
of a single chain can thus be expressed as
$\psi^{\scas{fjc}}=-k\,\theta \ln \, (p)$
where $\theta$ is the absolute temperature.
Depending on the range of stretching, either Gaussian or non-Gaussian
statistics are commonly applied in order to specify the particular
entropy changes upon deformation. %\\
For the classical Gaussian case for which
$p= p_0 \exp (-3/2 \,N \, r^2/L^2)$,
the free energy of an individual chain takes the following form
\beq
  \psi^{\scas{fjc}}_{\scas{gau}}
= \psi_0^{\scas{fjc}}
+  k \, \theta \, N \, \frac{3}{2} \, \frac{r^2}{L^2}
\eeq
where $\psi_0^{\scas{fjc}}$ is the value of the chain energy in the unperturbed state.
The single chain force
%$f^{\scas{fjc}} = \sca{d} \psi^{\scas{fjc}} / \sca{d} \lambda$
%dual to the stretch $r/L$
follows straightforwardly as
\beq
  f^{\scas{fjc}}_{\scas{gau}}
= k \, \theta \, N \, 3 \frac{r}{L}
\eeq
and is thus a linear function of the relative stretch $r/L$.
Alternatively, we could apply non-Gaussian statistics of inverse Langevin type
as introduced by Kuhn \& Gr\"un \cite{kuhn42}. With
$p =p_0 \exp(
-N {\mathcal{L}}^{-1} r/L
-N\ln({\mathcal{L}}^{-1}/\sinh({\mathcal{L}}^{-1})))$,
the strain energy $\psi^{\scas{fjc}}$ of the freely jointed chain can
be expressed as
\beq
  \psi^{\scas{fjc}}_{\scas{lan}}
= \psi_0^{\scas{fjc}}
+ k \, \theta \, N \left[ \frac{r}{L} \, {\mathcal{L}}^{-1}
+ \ln \left( \frac{{\mathcal{L}}^{-1}}{\sinh ({\mathcal{L}}^{-1})} \right) \right]
\eeq
where $\psi_0^{\scas{fjc}}$ is the value of the chain energy in the
unperturbed state and ${\mathcal{L}}^{-1}$ is the inverse Langevin function
with ${\mathcal{L}} \, (r/L) = \coth (r/L) - L/r$.
Accordingly,
\beq
  f^{\scas{fjc}}_{\scas{lan}}
= k \, \theta \, N {\mathcal{L}}^{-1}
\eeq
defines the force stretch relation for the freely jointed chain with inverse
Langevin statistics. Recall that the inverse Langevin function can be
evaluated by a Pad\'{e} approximation as
${\mathcal{L}}^{-1} \approx (3-r^2/L^2)/(1-r^2/L^2) \, r/L$,
as in Miehe et al. \cite{miehe04}.
At small stretches $r/L$ for which
${\mathcal{L}}^{-1} \approx 3 \, r/L$
the force of the inverse Langevin chain
$f^{\scas{fjc}}_{\scas{lan}} = k \, \theta \, N {\mathcal{L}}^{-1}$ thus obviously approaches
the force of the Gaussian chain
$f^{\scas{fjc}}_{\scas{gau}} = k \, \theta \, N \, 3 \, r/L$ .
%%%%%%%%%%%%%%%%%%%%%%%%%%%%%%%%%%%%%%%%%%%%%%%%%%%%%%%%%%%%%%%%%%%%%%%%
\subsection{Correlated chains: The wormlike chain model}\label{chain_wlc}
%%%%%%%%%%%%%%%%%%%%%%%%%%%%%%%%%%%%%%%%%%%%%%%%%%%%%%%%%%%%%%%%%%%%%%%%
The distinguishing feature of the wormlike chain or Kratky-Porod
model chain is the continuity of the direction of its contour in space, see
Kratky \& Porod \cite{kratky49} or Flory \cite{flory69}.
As such, it is characterized through a
smooth curvature whose direction changes randomly but in a continuous manner.
This property is essentially reflected through a second parameter besides the
contour length $L=N\,l$, namely the persistence length $A$.
The persistence length
can be understood as the sum of the average projection of all bonds onto the direction
of the first bond, as sketched in figure \ref{chains_kinem}, right.
Varying between $l \le A \le L$, the persistence length
is thus a particular measure of stiffness, see also
Landau \& Lifshitz \cite{landau51}. Accordingly,
the persistence length of the uncorrelated
freely jointed chain is equal to the length of the first bond $A=l$
while the persistence length of an infinitely stiff chain with almost beam-like
properties is equal to its contour length $A=L$.
The strain energy $\psi^{\scas{wlc}}$ of the wormlike chain model
\beq
  \psi^{\scas{wlc}}
= \psi_0^{\scas{wlc}}
+ \frac{k \, \theta \, L}{4 A}
  \left[ 2 \frac{r^2}{L^2} + \frac{1}{[1-r/L]} - \frac{r}{L} \right]
\label{psi_wlc}
\eeq
can be derived straightforwardly by integrating the
force stretch relation for a wormlike chain
\beq
  f^{\scas{wlc}}
= \frac{k \, \theta}{4 A}
  \left[ 4 \frac{r}{L} + \frac{1}{[1-r/L]^2} - \, 1 \, \right]
\eeq
as originally suggested for the DNA double helix by
Marko \& Siggia \cite{marko95} and
Bustamante et al. \cite{bustamante03}
and applied for the collagen triple helix by Bischoff et al. \cite{bischoff02}.
Again,
$\psi_0^{\scas{wlc}}$ is the value of the chain energy in the unperturbed state.
%%%%%%%%%%%%%%%%%%%%%%%%%%%%%%%%%%%%%%%%%%%%%%%%%%%%%%%%%%%%%%%%%%%%%%%%
\subsection{Example: Comparison of different chain models}
\begin{figure}[h]
\unitlength1.0mm
\begin{picture}(139.0,58.0)
\put( -1.0,-2.0){\includegraphics[width=
70.mm,angle=0]{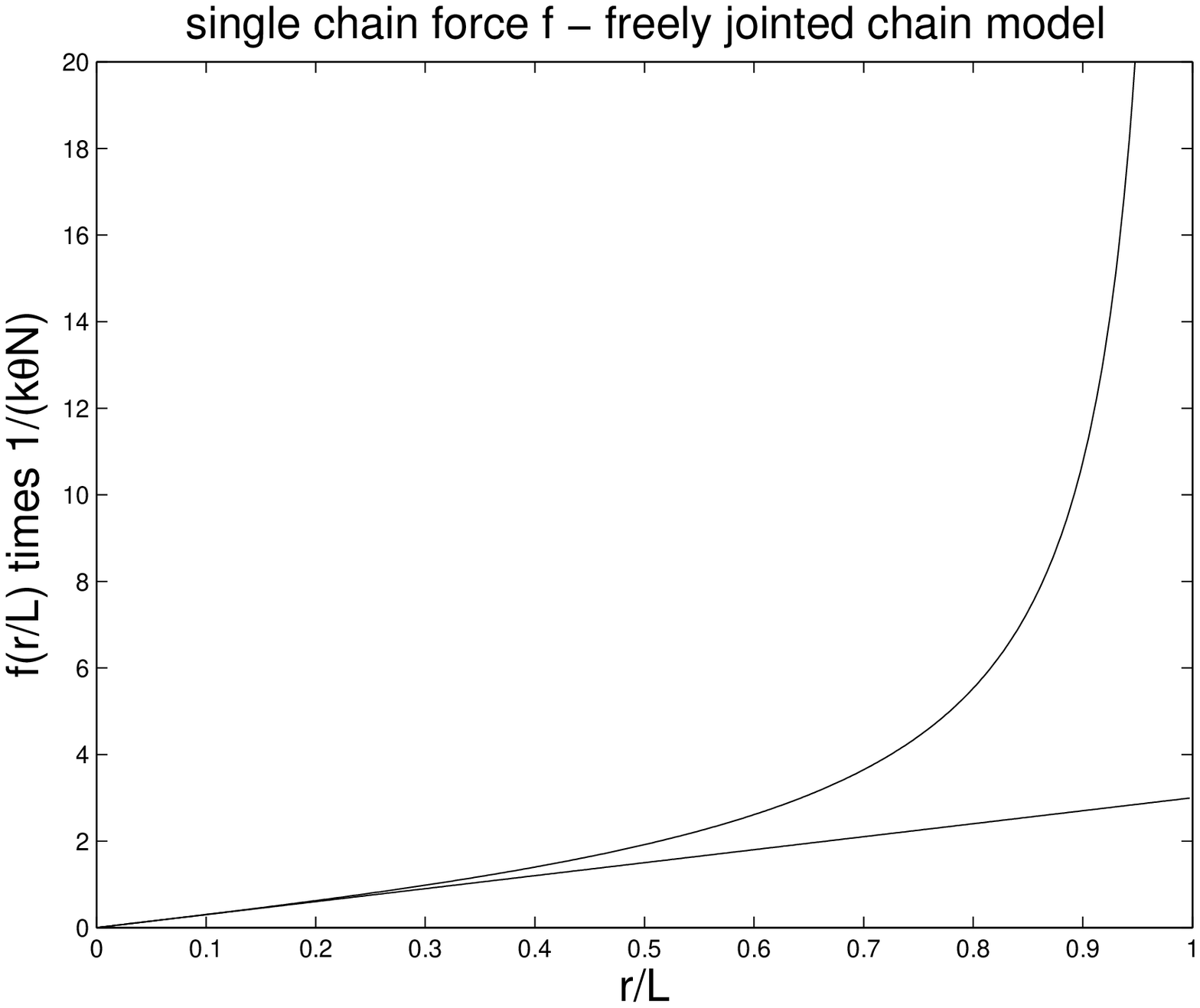}} \put(
7.5,31.0){\includegraphics[width= 25.mm,angle=0]{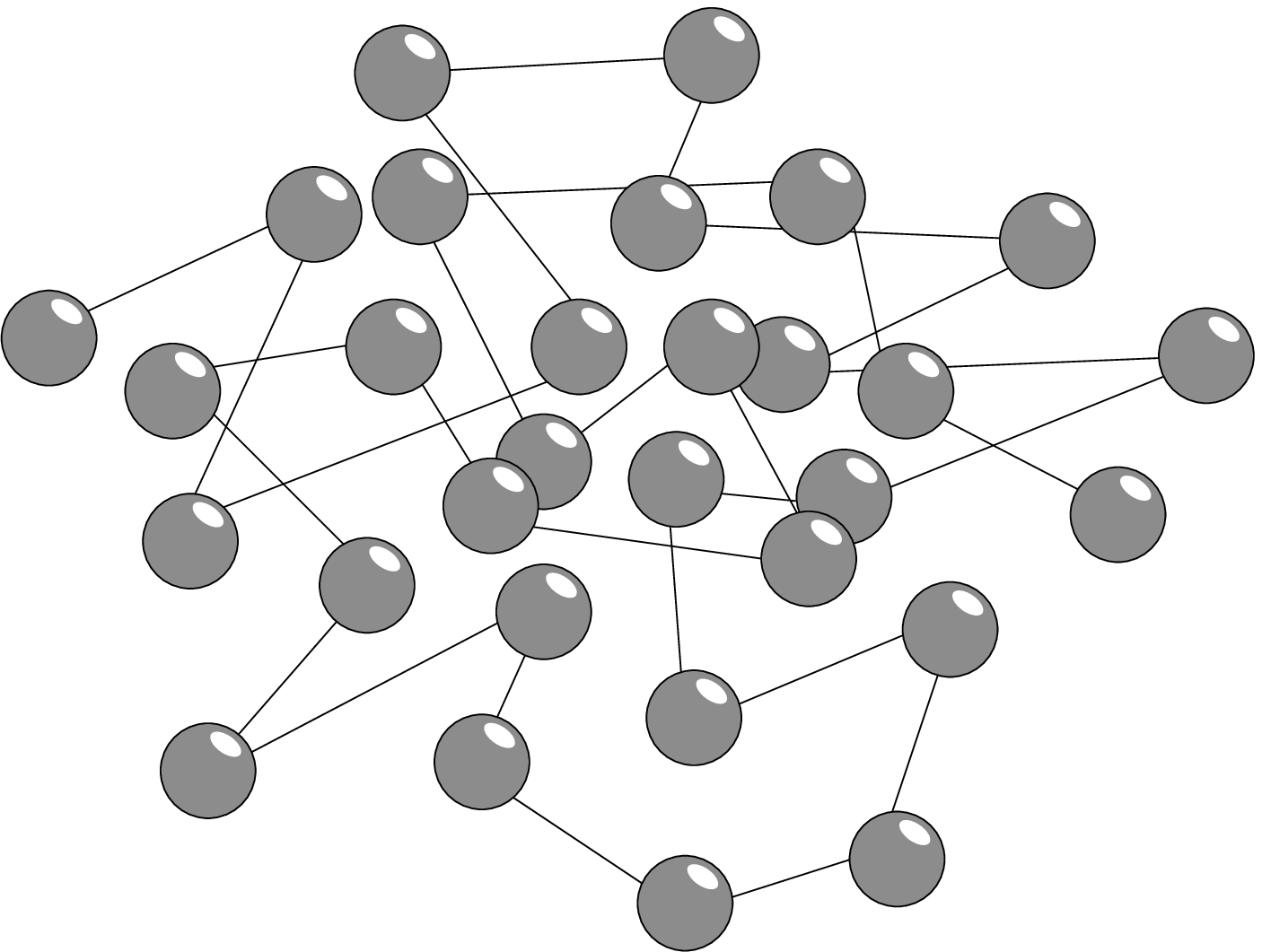}} \put(
69.5,-2.0){\includegraphics[width= 70.mm,angle=0]{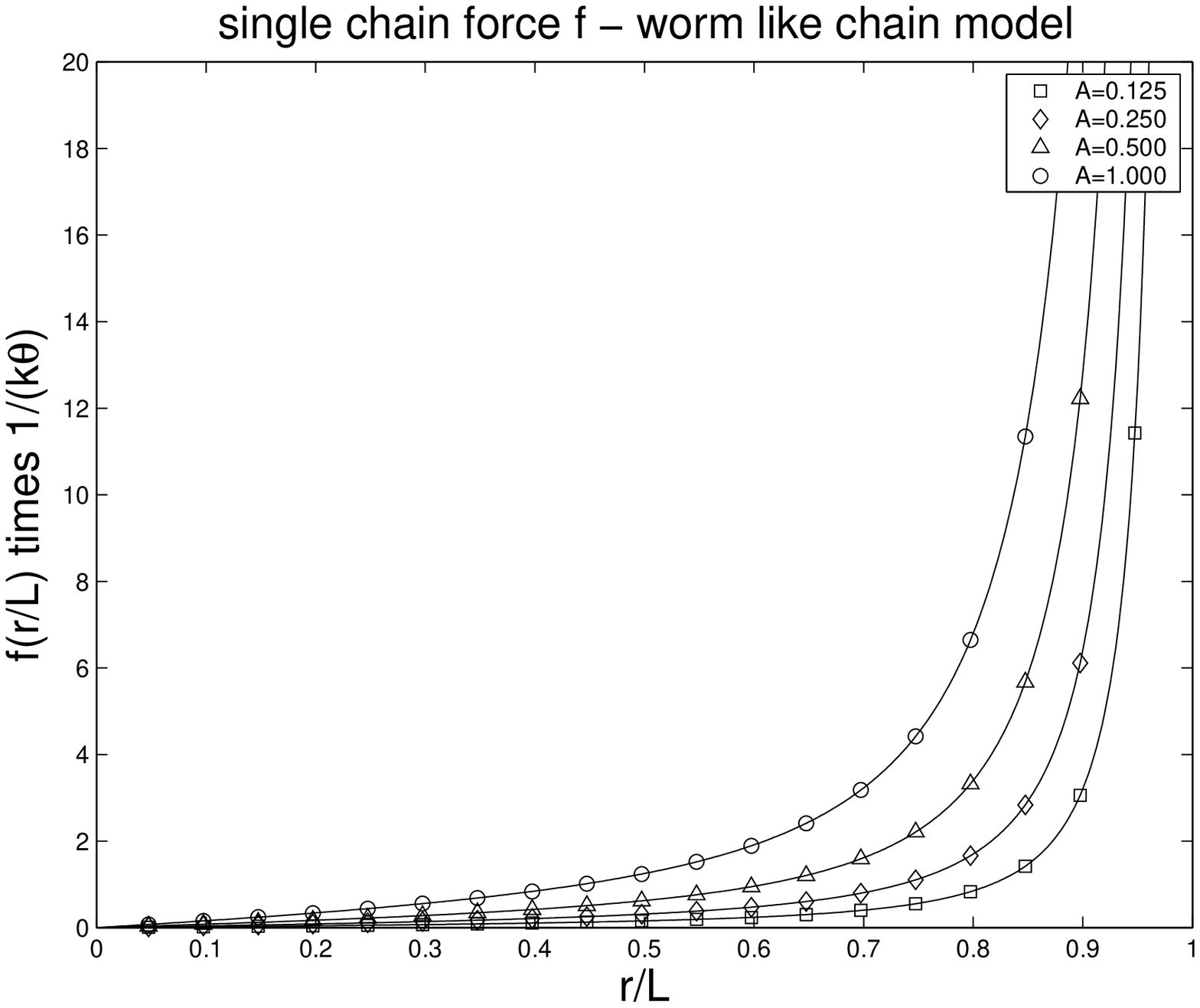}}
\put( 77.0,34.5){\includegraphics[width=
32.mm,angle=8]{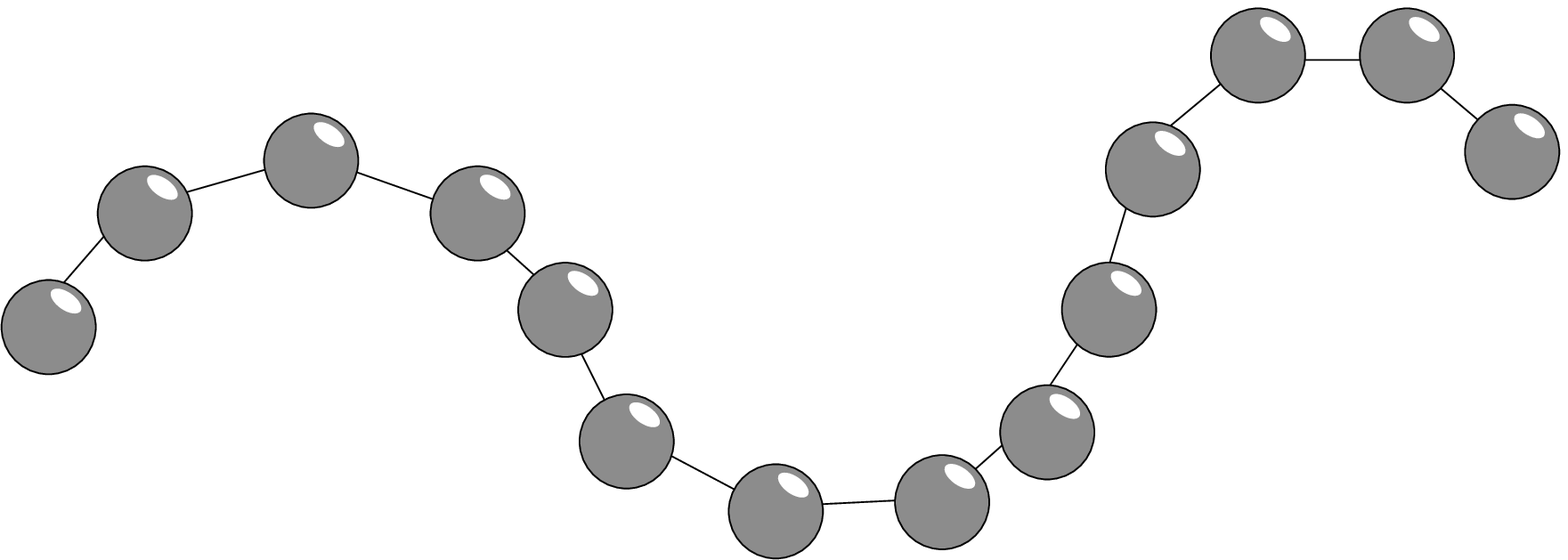}}
\end{picture}
\caption{Force vs. elongation response of individual chain models}
\label{forces_models}
\end{figure}

%%%%%%%%%%%%%%%%%%%%%%%%%%%%%%%%%%%%%%%%%%%%%%%%%%%%%%%%%%%%%%%%%%%%%%%%
To illustrate the fundamental differences between the freely jointed chain model
and the wormlike chain model, we plot the different force elongation curves
for two individual model chains. Figure \ref{forces_models}, left,
depicts the force elongation behavior of both, a Gaussian and an inverse Langevin
freely jointed chain with the force $f^{\scas{fjc}}$ being
scaled by the factor $1/[k \, \theta \, N]$. The two curves clearly monitor
the deviation of Gaussian and the inverse Langevin statistics in the large
strain regime, for which the linear force elongation behavior of
Gaussian statistics is no longer appropriate. The inverse Langevin
freely jointed chain model, however, nicely captures
the characteristic locking behavior close to the locking stretch $r=L$.
In the low strain region, the chain shows nearly no resistance to loading
while close to its full extension, the chain stiffness increases considerably.
%%%%%%%%%%%%%%%%%%%%%%%%%%%%%%%%%%%%%%%%%%%%%%%%%%%%%%%%%%%%%%%%%%%%%%%%
\\
%%%%%%%%%%%%%%%%%%%%%%%%%%%%%%%%%%%%%%%%%%%%%%%%%%%%%%%%%%%%%%%%%%%%%%%%%
Next, we elaborate the force elongation response of a single wormlike chain
model. Figure \ref{forces_models}, right, shows the chain force
$f^{\scas{wlc}}$ scaled by $1/[k\,\theta]$ for varying chain stretches $r/L$
at different persistence lengths, $A$, varying from $A=0.125$, right curve,
to $A=1.000$, left curve.
Again, the characteristic locking behavior is nicely
captured by the model as $r/L$ approaches one.
However, in contrast to the single parameter freely jointed chain
model, the wormlike chain model offers the additional freedom of a second parameter, namely
the persistence length. With this second parameter,
the wormlike chain model is not only able fit the locking stretch
but also to capture the shape of the force elongation curve appropriately.
For larger values of of the
persistence length indicating
initially stiffer chains, the locking response, i.e. the behavior in the
$r\,/\,L \rightarrow 1$ regime, is much smoother.
For smaller values of $A$, the strong locking behavior
of the uncorrelated freely jointed chain
characterized through a steep slope of the force elongation curve
can be captured.\\
%$A=0.25$, $L=1$
%%%%%%%%%%%%%%%%%%%%%%%%%%%%%%%%%%%%%%%%%%%%%%%%%%%%%%%%%%%%%%%%%%%%%%%%
While the freely jointed chain model of figure \ref{forces_models}, left,
shows a pronounced locking behavior,
the stiffness of the wormlike chain model of figure \ref{forces_models}, right,
increases gradually as the locking stretch is approached.
For the densely packed collagen fibrils considered in the
present work, the two parameter
wormlike chain model is believed to represent the real
material behavior more accurately than the
single parameter freely jointed chain model which
might be better suited for randomly oriented polymer chains in rubber.
In particular due to the additional freedom introduced by the
persistence length as a second parameter, we shall thus focus
on the wormlike chain model for the collagen fibrils in the sequel.

%%%%%%%%%%%%%%%%%%%%%%%%%%%%%%%%%%%%%%%%%%%%%%%%%%%%%%%%%%%%%%%%%%%%%%%%
\section{Mechanics of the chain network}\label{homog}
%%%%%%%%%%%%%%%%%%%%%%%%%%%%%%%%%%%%%%%%%%%%%%%%%%%%%%%%%%%%%%%%%%%%%%%%
To incorporate the individual chain statistics into an overall constitutive
description, we apply an eight chain representation of the underlying cooperative
%%%%%%%%%%%%%%%%%%%%%%%%%%%%%%%%%%%%%%%%%%%%%%%%%%%%%%%%%%%%%%%%%%%%%%%%
\begin{figure}[t]
%\begin{center}
\unitlength1.0mm
\begin{picture}(139.0,64.0)
\put(0.0, 0.0){\includegraphics[width=139mm,angle=0]{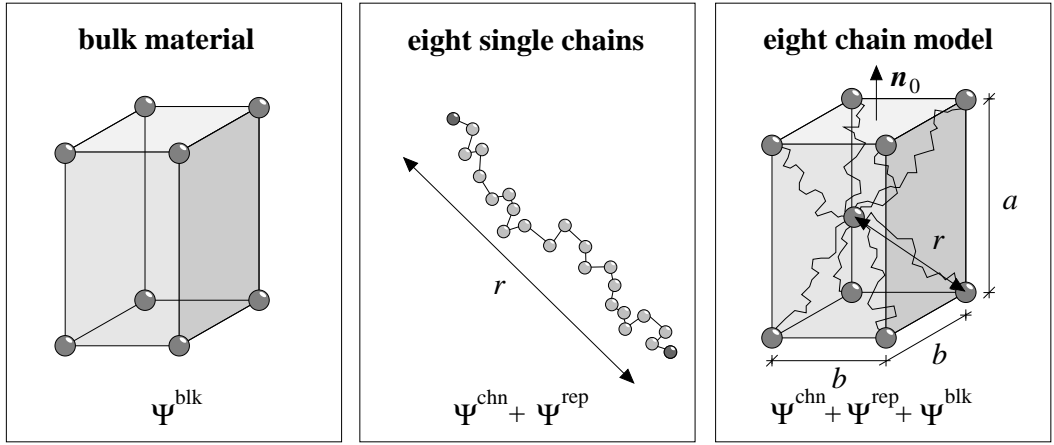}}
\end{picture}
%\end{center}
\caption{Kinematics of transversely isotropic eight chain network model}
\label{eight_ortho}
\end{figure}
%%%%%%%%%%%%%%%%%%%%%%%%%%%%%%%%%%%%%%%%%%%%%%%%%%%%%%%%%%%%%%%%%%%%%%%
macromolecular network structure. These eight chains are embedded
in a transversely isotropic unit cell with initial
cell dimensions $a$ and $b$. They essentially
link at its center and extend to the eight individual corners
as in figure \ref{eight_ortho}.
While the end-to-end length in the undeformed configuration is obviously
given as $r_0 = \sqrt{ a^2 + b^2 + b^2 }/2$, we
assume that the end-to-end length in the deformed configuration
can be expressed as
\beq
r = \sqrt{ I_4 \, a^2 +  [I_1-I_4] \, b^2}\,/\,2
\eeq
in terms of the first and fourth invariant $I_1$ and $I_4$.
%Accoringly, the
%out-of-plane stretch ${\lambda_{a}}$ is represented in an {\it affine} way while the
%isotropic in-plane stretch is captured in a {\it non-affine} manner.
The relevant
invariants and their derivatives are given in the following form.
\beq
\begin{array}{r@{\hspace*{0.2cm}}c@{\hspace*{0.2cm}}
              l@{\hspace*{1.0cm}}
              r@{\hspace*{0.2cm}}c@{\hspace*{0.2cm}}
              l@{\hspace*{1.0cm}}
              r@{\hspace*{0.2cm}}c@{\hspace*{0.2cm}}l}
I_1 &=& \ten{G}^{\scas{-1}} : \ten{C}&
I_3 &=& \det{} (\ten{C})&
I_4 &=& \ten{N}_0 : \ten{C}  \\
\sca{d}_{\tens{C}} I_1 &=& \ten{G}^{\scas{-1}} &
\sca{d}_{\tens{C}} I_3 &=& I_3 \, \ten{C}^{\scas{-t}} &
\sca{d}_{\tens{C}} I_4 &=& \ten{N}_0
\end{array}
\eeq
The first invariant $I_1$ can either be expressed as the trace of the covariant
Cauchy Green strain tensor $\ten{C}$ as $I_1=\ten{G}^{\scas{-1}}:\ten{C}$
or of the contravariant finger tensor $\ten{b}$ as $I_1=\ten{g}:\ten{b}$.
Thereby, the Cauchy Green
tensor $\ten{C}$ is defined as the pull back of the covariant spatial
metric $\ten{g}$
whereas the finger tensor $\ten{b}$ is the
push forward of the contravariant material metric $\ten{G}^{\scas{-1}}$.
\beq
\ten{C}=\ten{F}^{\scas{t}}\cdot \ten{g} \cdot \ten{F}
\qquad \quad
\ten{b}=\ten{F}\cdot \ten{G}^{\scas{-1}} \cdot \ten{F}^{\scas{t}}
\eeq
Herein, $\ten{F}$ denotes the deformation gradient as
$\ten{F}=\nabla_{\vecs{X}} \vec{\varphi}$ with
$\vec{\varphi}$ being the deformation map between the undeformed and the
deformed configuration.
The determinant of either $\ten{C}$ or $\ten{b}$
defines the third invariant $I_3$ which is thus identical to the Jacobian
squared $I_3=J^2$, whereby $J = \det{} (\ten{F})$.
The fourth invariant $I_4=\lambda_a^2$
essentially represents the square of the stretch
$\lambda_a$ along the $a$ direction.
As such, $I_4$ can be expressed as the scalar product of the Cauchy Green strain
$\ten{C}$ with the structural tensor
$\ten{N}_0=\vec{n}_0 \otimes \vec{n}_0$, i.e. the dyadic product of the
normal vectors $\vec{n}_0$ in the undeformed configuration.
Note that while the stretch $\lambda_a$
in the out-of-plane direction is captured in an
{\it affine} way through the $I_4$ term, the in-plane stretch
$\lambda_b$ is obviously not uniquely defined. It is thus represented in a
{\it non-affine} manner through the $[I_1-I_4]$ term.
The in-plane term $[I_1-I_4]$ obviously results from the scalar product of
$\ten{C}$ with the remaining contribution
$\ten{G}^{-1}-\ten{N}_0 = \ten{G}^{-1} -\vec{n}_0 \otimes \vec{n}_0$.
Recall that in the undeformed configuration
$\lambda_a = \lambda_b = 1$,
$I_1 = 3$,
$I_4 = 1$ and thus
$r_0 = \sqrt{ a^2 + b^2 + b^2 }/2$.
The overall energy $\Psi$ of the transversely isotropic eight chain unit cell
is assumed to consist of three contributions
\beq
\Psi = \Psi^{\scas{blk}} + \Psi^{\scas{chn}} + \Psi^{\scas{rep}}
\eeq
as illustrated in figure \ref{eight_ortho}.
The first term $\Psi^{\scas{blk}} (I_1, I_3)$ captures the effect of
bulk incompressibility, e.g. due to a surrounding
liquid solvent, and is thus of isotropic nature,
see Garikipati et al. \cite{garikipati04}.
The second term $\Psi^{\scas{chn}} (I_1,I_4)$
reflects the effective assembly of the individual eight chain
energies $\psi^{\scas{chn}}$. Accordingly,
$\Psi^{\scas{chn}}= \gamma^{\scas{chn}}\; \psi^{\scas{chn}}$
with $\gamma^{\scas{chn}}$ denoting the chain density per unit cell.
The repulsive term $\Psi^{\scas{rep}}(I_1,I_4)$ accounts for a
stress-free reference configuration and prevents the material from collapsing.
The second and third terms essentially depend on the key phenomenological kinematic
variable of a single chain, the current end-to-end length $r (I_1,I_4)$.
For the wormlike chain model based on the free energy
$\psi^{\scas{chn}}=\psi^{\scas{wlc}}$
according to equation (\ref{psi_wlc}),
the individual energy terms take the following expressions,
\beq
\begin{array}{l@{\hspace*{0.05cm}}l@{\hspace*{0.05cm}}
              l@{\hspace*{0.05cm}}l}
 \D{\Psi^{\scas{blk}}}
&\D{=}
&\D{\quad \gamma^{\scas{blk}}}
&\D{\left[ \;
       I_1 - 3 \,
    +  \frac{1}{\beta} \, [\, I_3^{-\beta} -1 \,] \,
    \right]}\\
 \D{\Psi^{\scas{chn}}}
&\D{=}
&\D{\frac{\gamma^{\scas{chn}} k \theta L}{4 \, A}}
&\D{\left[
     2 \frac{r^2}{L^2}
    +\frac{1}{[\,1- r/L\,]}
    -\frac{r}{L}
    \right]}\\
 \D{\Psi^{\scas{rep}}}
&\D{=-}
&\D{\frac{\gamma^{\scas{chn}} k \theta}{4 \, A}}
&\D{\left[
     \frac{1}{L}
    +\frac{1}{4 r_0 [1- r_0/L]^2}
    -\frac{1}{4 r_0}
    \right]
    \bar{\Psi}^{\scas{rep}}}
\end{array}
\label{psi_terms}
\eeq
where we have introduced the following abbreviation for the repulsive
weighting factor $\bar{\Psi}^{\scas{rep}}$.
\beq
 \bar{\Psi}^{\scas{rep}}
=\ln \, (I_4^{[a^2-b^2]/2})
+\frac{3}{2} \ln \, (I_1^{b^2})
\label{psi_rep}
\eeq
The parameter set of the model is thus restricted to
the chain density $\gamma^{\scas{blk}}$,
the two wormlike chain parameters,
i.e. the persistence length $A$ and the
contour length $L$,
the cell dimensions $a$ and $b$
and the two bulk parameters $\gamma^{\scas{blk}}$ and $\beta$.
The above introduced free energy $\Psi$ defines the
Kirchhoff stress
\nolinebreak[4]$\ten{\tau}$
\beq
\ten{\tau} = \ten{\tau}^{\scas{blk}} + \ten{\tau}^{\scas{chn}} + \ten{\tau}^{\scas{rep}}
\eeq
as the contravariant push forward of the second Piola Kirchhoff stress
$2\, \sca{d}_{\tens{C}} \, \Psi$
as
$\ten{\tau}=\ten{F}\cdot 2 \, \sca{d}_{\tens{C}} \, \Psi \cdot\ten{F}^{\scas{t}}$.
%$\ten{S}=\sca{d}_{\tens{C}} \, \Psi$
The individual stress contributions follow straightforwardly from
the corresponding energy terms introduced in equations
\nolinebreak[4](\ref{psi_terms})
\beq
\begin{array}{ @{\hspace*{0.0cm}}l@{\hspace*{0.05cm}}
              l@{\hspace*{0.05cm}}r@{\hspace*{0.05cm}}
              c@{\hspace*{0.05cm}}l}
 \D{\ten{\tau}^{\scas{blk}}}
&\D{=}
&\D{\gamma^{\scas{blk}} \;\;}
&\D{\left[ \;2 \, \ten{b}
      - 2 \, I_3^{-\beta} \ten{g}^{\scas{-1}} \right]}\\[1.ex]
 \D{\ten{\tau}^{\scas{chn}}}
&\D{=}
&\D{\frac{\gamma^{\scas{chn}} k \theta}{4 A}}
&\D{\left[
     \frac{1}{L}
    +\frac{1}{4 \, r \; \, [1- r \, /L]^2 \;}
    -\frac{1}{4 \; r \; }
    \right]}
&\D{\bar{\ten{\tau}}^{\scas{chn}}}\\
 \D{\ten{\tau}^{\scas{rep}}}
&\D{=-}
&\D{\frac{\gamma^{\scas{chn}} k \theta}{4 A}}
&\D{\left[
     \frac{1}{L}
    +\frac{1}{4 \, r_0 [1- r_0/L]^2}
    -\frac{1}{4 \, r_0}
    \right]}
&\D{\bar{\ten{\tau}}^{\scas{rep}}}
\label{tau_terms}
\end{array}
\eeq
where $\ten{g}^{\scas{-1}}$ denotes the contravariant spatial metric.
In the above equations,
we have made use of the following abbreviations for the bases of the
chain stress and of the repulsive stress.
\beq
\begin{array}{l@{\hspace*{0.0cm}}l@{\hspace*{0.1cm}}
              c@{\hspace*{0.1cm}}c@{\hspace*{0.1cm}}
              c@{\hspace*{0.1cm}}
              r@{\hspace*{0.1cm}}c@{\hspace*{0.1cm}}l}
 \D{\bar{\ten{\tau}}^{\scas{chn}}}
&\D{=}
&\D{}
&\D{[\, a^2 - b^2 \,]}
&\D{\ten{N}}
&\D{+}
&\D{}
&\D{b^2 \, \ten{b}}\\
 \D{\bar{\ten{\tau}}^{\scas{rep}}}
&\D{=}
&\D{\frac{1}{I_4}}
&\D{[\, a^2 - b^2 \,]}
&\D{\ten{N}}
&\D{+}
&\D{\frac{3}{I_1}}
&\D{b^2 \, \ten{b}}.
\label{abbr_stress}
\end{array}
\eeq
Here, $\ten{N}=\ten{F} \cdot \ten{N}_0 \cdot \ten{F}^{\scas{t}}$
denotes the push forward of the structural tensor $\ten{N}_0$.
As such, it can be expressed as
$\ten{N} = \bar{\vec{n}} \otimes \bar{\vec{n}}$
in terms of the cell orientation of the deformed configuration
$\bar{\vec{n}}_a = \ten{F} \cdot \vec{n}_a^0$.
In equation (\ref{abbr_stress}), we have introduced the abbreviation
$\bar{\ten{\tau}}^{\scas{chn}}
=\ten{F} \cdot 8\,r\,\sca{d}_{\tens{C}} r \cdot \ten{F}^{\scas{t}}$
for the derivative of the end-to-end length $r$.
Since we assume a stress-free initial state at $r=r_0$, we conclude that
$\ten{\tau}^{\scas{rep}} (r_0) \doteq -\ten{\tau}^{\scas{chn}}(r_0)$
and thus
$\bar{\ten{\tau}}^{\scas{rep}}(r_0) \doteq \bar{\ten{\tau}}^{\scas{chn}}(r_0)$.
Accordingly, the repulsive energy contribution
$\bar{\Psi}^{\scas{rep}}$
introduced in equation (\ref{psi_terms})
has been constructed in such a way that
$\bar{\ten{\tau}}^{\scas{rep}}
=\ten{F}\cdot 2 \, \sca{d}_{\tens{C}} \, \bar{\Psi}^{\scas{rep}}
\cdot\ten{F}^{\scas{t}}$.
With the help of the above equations,
the spatial Kirchhoff tangent
$\ten{\mathcal{C}}$ %=4 \partial_{\tens{g}\otimes\ten{g}} \Psi$
relating the Lie derivative of the Kirchhoff stress
$\sca{L}_t \ten{\tau}$
to the Lie derivative of the covariant spatial metric
$\sca{L}_t \ten{g}$ as
$\sca{L}_t \ten{\tau}=\ten{\mathcal{C}}:\sca{L}_t \ten{g}/2$ with
\beq
  \ten{\mathcal{C}}
= \ten{\mathcal{C}}^{\scas{blk}}
+ \ten{\mathcal{C}}^{\scas{chn}}
+ \ten{\mathcal{C}}^{\scas{rep}}
\eeq
is defined through the contravariant push forward of the material tangent
$4 \, \sca{d}_{\tens{C}\otimes\tens{C}} \Psi$
as
$\ten{\mathcal{C}}=
[\ten{F} \overline{\otimes} \ten{F}]
: 4 \, \sca{d}_{\tens{C}\otimes\tens{C}} \Psi:
[\ten{F}^{\scas{t}} \overline{\otimes} \ten{F}^{\scas{t}}]$. Its individual
contributions take the following representation.
\beq
\begin{array}{ @{\hspace*{0.0cm}}
              l@{\hspace*{0.0cm}}l@{\hspace*{0.0cm}}
              l@{\hspace*{0.0cm}}l@{\hspace*{0.0cm}}c}
 \D{\ten{\mathcal{C}}^{\scas{blk}}}
&\D{=}
&\D{\quad  4 \, \gamma^{\scas{blk}}}
&\D{\left[ I_3^{-\beta} \, \ten{\mathcal{I}}
      +  \beta \,I_3^{- \beta} \, \ten{g}^{\scas{-1}} \otimes \ten{g}^{\scas{-1}} \right]} \\
 \D{\ten{\mathcal{C}}^{\scas{chn}}}
&\D{=}
&\D{\frac{\gamma^{\scas{chn}} k \theta}{64 A r^3}}
&\D{\left[
      1
    -\frac{1}{1 - [1- r/L]^2}
    +\frac{2 \, r}{L \, [1- r/L]^3}
    \right]}
&\D{\bar{\ten{\mathcal{C}}}^{\scas{chn}}} \\
 \D{\ten{\mathcal{C}}^{\scas{rep}}}
&\D{=}
&\D{-\frac{\gamma^{\scas{chn}} k \theta}{4 A}}
&\D{\left[
     \frac{1}{L}
    +\frac{1}{4 \, r_0 [1- r_0/L]^2}
    -\frac{1}{4 \, r_0}
    \right]}
&\D{\bar{\ten{\mathcal{C}}}^{\scas{rep}}}
\end{array}
\label{c_terms}
\eeq
The fourth order basis of the chain term
$\bar{\ten{\mathcal{C}}}^{\scas{chn}}
=\bar{\ten{\tau}}^{\scas{chn}} \, \otimes \, \bar{\ten{\tau}}^{\scas{chn}}$
and of the
repulsive term
$\bar{\ten{\mathcal{C}}}^{\scas{rep}}
=\ten{F}\cdot2\,\sca{d}_{\tens{C}} \bar{\ten{\tau}}^{\scas{rep}}
 \cdot \ten{F}^{\scas{t}}$
can be expressed as follows.
\beq
\begin{array}{lcl}
 \D{\bar{\ten{\mathcal{C}}}^{\scas{chn}}}
&\D{=}
&\D{[ \,[ a^2 - b^2 ] \ten{N} + b^2 \, \ten{b} \,]
    \otimes
    [ \,[ a^2 - b^2 ] \ten{N} + b^2 \, \ten{b} \,]}\\[1.ex]
 \D{\bar{\ten{\mathcal{C}}}^{\scas{rep}}}
&\D{=}
&\D{-\frac{2}{I_4^2} \, [a^2-b^2] \, \ten{N} \otimes \ten{N}
    -\frac{6}{I_1^2} \,      b^2  \, \ten{b}   \otimes \ten{b}}
\end{array}
\eeq
In equation (\ref{c_terms}), $\ten{\mathcal{I}}$ denotes the
fourth order identity which can be expressed as
$\ten{\mathcal{I}}
=[ \,\ten{g}^{\scas{-1}} \,  \overline{\otimes} \, \ten{g}^{\scas{-1}}
+    \ten{g}^{\scas{-1}} \, \underline{\otimes} \, \ten{g}^{\scas{-1}} \, ]\,/\, 2$.
Here, we have applied the abbreviations
$\overline{\otimes}$ and $\underline{\otimes}$
for the non--standard dyadic products according to the following
component-wise definitions
$\{ \bullet  \overline{\otimes}   \circ\}_{ijkl}
=\{ \bullet \}_{ik}    \otimes \{ \circ\}_{jl}$ and
$\{ \bullet \underline{\otimes}   \circ\}_{ijkl}
=\{ \bullet \}_{il}    \otimes \{ \circ\}_{jk}$.
Recall that transverse isotropy is basically defined through the normal
$\vec{n}_0$ of the
preferred material direction. This normal will be used later on to specify
biological remodeling as it is assumed to represent the orientation of
collagen fibers within a tissue.
%%%%%%%%%%%%%%%%%%%%%%%%%%%%%%%%%%%%%%%%%%%%%%%%%%%%%%%%%%%%%%%%%%%%%%%%
\begin{remark}[Special case of an isotropic network model]
%%%%%%%%%%%%%%%%%%%%%%%%%%%%%%%%%%%%%%%%%%%%%%%%%%%%%%%%%%%%%%%%%%%%%%%%
$\quad$
The classical non-affine isotropic eight chain model of
Arruda \& Boyce \cite{arruda93,boyce96,boyce00}
which was originally introduced in the context of rubber elasticity
can be understood as a special case of the present framework. Its undeformed
unit cell represents a cube with $a=b$.
The chain extension thus reduces to a
function of the the first strain invariant $I_1$ such that
$r=\sqrt{I_1} \,a / 2$.
For the isotropic eight chain model,
the undeformed reference configuration is characterized through
$\lambda_a = \lambda_b = 1$,
$I_1 = 3$ and thus
$r_0 = \sqrt{3} \, a / 2$.
The corresponding repulsive term
$ \bar{\Psi}^{\scas{rep}} = {3}/{2} \, \ln \, (I_1^{a^2})$
introduces the stress contributions
$\bar{\ten{\tau}}^{\scas{chn}}=a^2 \, \ten{b}$ and
$\bar{\ten{\tau}}^{\scas{chn}}= 3/I_1 \, a^2 \, \ten{b}$.
Accordingly, the bases for the chain and the repulsive tangent term
which follow as
$ \bar{\ten{\mathcal{C}}}^{\scas{chn}}
= a^4 \, \ten{b} \, \otimes \, \ten{b}$
and
$ \bar{\ten{\mathcal{C}}}^{\scas{rep}}
=-{6 \, a^2}/{I_1^2} \, \ten{b} \, \otimes \, \ten{b}$
clearly reflect the isotropic nature of this particular specification of
the model.
\end{remark}
%%%%%%%%%%%%%%%%%%%%%%%%%%%%%%%%%%%%%%%%%%%%%%%%%%%%%%%%%%%%%%%%%%%%%%%%
\begin{remark}[Special case of a transversely isotropic model]
%%%%%%%%%%%%%%%%%%%%%%%%%%%%%%%%%%%%%%%%%%%%%%%%%%%%%%%%%%%%%%%%%%%%%%%%
$\quad$
Another special case of the transversely isotropic chain network model follows
from assuming a degenerated unit cell for which the in-plane dimension
tends to zero as $b = 0$.
This affine chain model is based on the introduction of chains which are all
oriented in a single direction $\vec{n}_0$.
%The end-to-end length of the deformed configuration is then exclusively
%given in terms of the affine projection of the overall strains
%onto the relevant cell direction $a$ as
%$r = \sqrt{I_4} \, a \,/\, 2$ such that
%$r_0 = a \,/\, 2$.
The deformed end-to-end length $r=\sqrt{I_4} \, a \,/\,2$ thus
degenerates to a mere function of
the fourth invariant $I_4$, or rather of the stretch $\lambda_a$,
and does no longer incorporate cross-linking network effects.
In the undeformed case with $I_4=1$, the end-to-end length is $r_0=a\,/\,2$.
The repulsive term of equation (\ref{psi_rep}) thus degenerates to
$\bar{\Psi}^{\scas{rep}} = \ln \,(I_4^{a^4/2})$. The corresponding bases of the
stress and tangent contributions are then given exclusively in terms of the
structural tensor $\ten{N}$ as
$\bar{\ten{\tau}}^{\scas{chn}}=a^2 \, \ten{N}$,
$\bar{\ten{\tau}}^{\scas{rep}}=1\,/\,I_4 \, a^2 \, \ten{N}$,
$\bar{\ten{\mathcal{C}}}^{\scas{chn}}=a^4 \, \ten{N} \otimes \ten{N}$ and
$\bar{\ten{\mathcal{C}}}^{\scas{rep}}=- 2\,/\, I_4 \, a^2 \, \ten{N} \otimes
\ten{N}$.
Obviously, this specific representation of the
model, which has been termed ``decoupled reinforcement model'' by
Merodio and Ogden (2002),(2003) does not include
the characteristic cross-link effects of the network structure.
\end{remark}
%%%%%%%%%%%%%%%%%%%%%%%%%%%%%%%%%%%%%%%%%%%%%%%%%%%%%%%%%%%%%%%%%%%%%%%%
\subsection{Example: Influence of anisotropy}
%%%%%%%%%%%%%%%%%%%%%%%%%%%%%%%%%%%%%%%%%%%%%%%%%%%%%%%%%%%%%%%%%%%%%%%%
In the present theory, transverse isotropy is
represented by the particular initial cell orientation $\vec{n}_0$ and by
the unit cell dimensions $a$ and $b$.
The influence of both will be studied in the sequel for the illustrative
homogeneous load case of uniaxial tension. The eight chains of the model are of
wormlike chain type as introduced in section \ref{chain_wlc}.
In anticipation of later examples,
the contour length $L$ and the persistence length $A$ have been chosen as
$L=2.125$ and $A=1.82$. The chain density is taken as
$\gamma^{\scas{chn}}=7 \times 10^{21}$.
The bulk material of the unit cell is defined by the
parameters $\gamma^{\scas{blk}}=100$ and $\beta=4.5$.
%%%%%%%%%%%%%%%%%%%%%%%%%%%%%%%%%%%%%%%%%%%%%%%%%%%%%%%%%%%%%%%%%%%%%%%%
\begin{figure}[h]
\unitlength1.0mm
\begin{picture}(139.0,58.0)
\put( -1.0,-2.0){\includegraphics[width= 70.mm,angle=0]{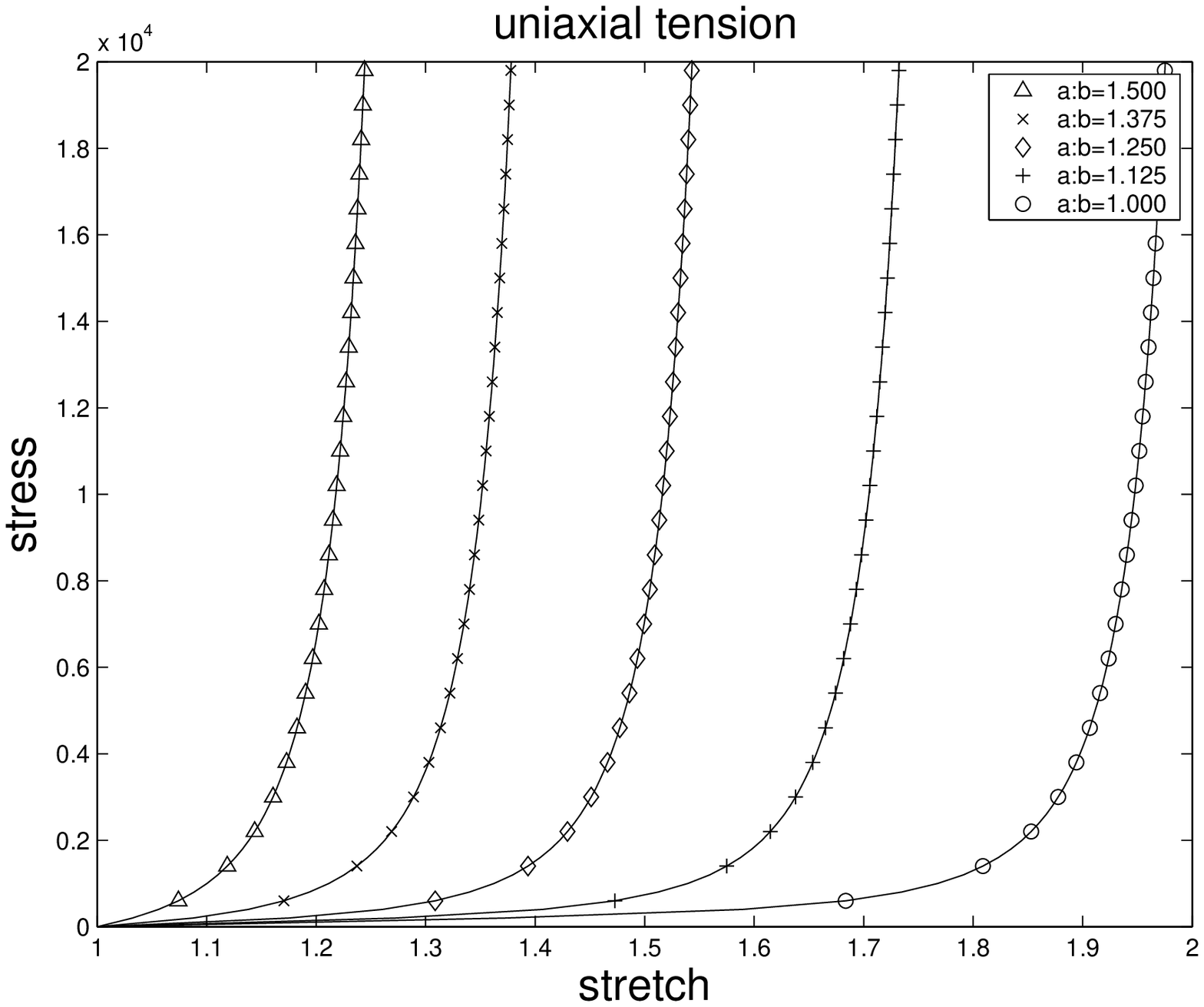}}
\put(  6.4,21.5){\includegraphics[width=
20.mm,angle=0]{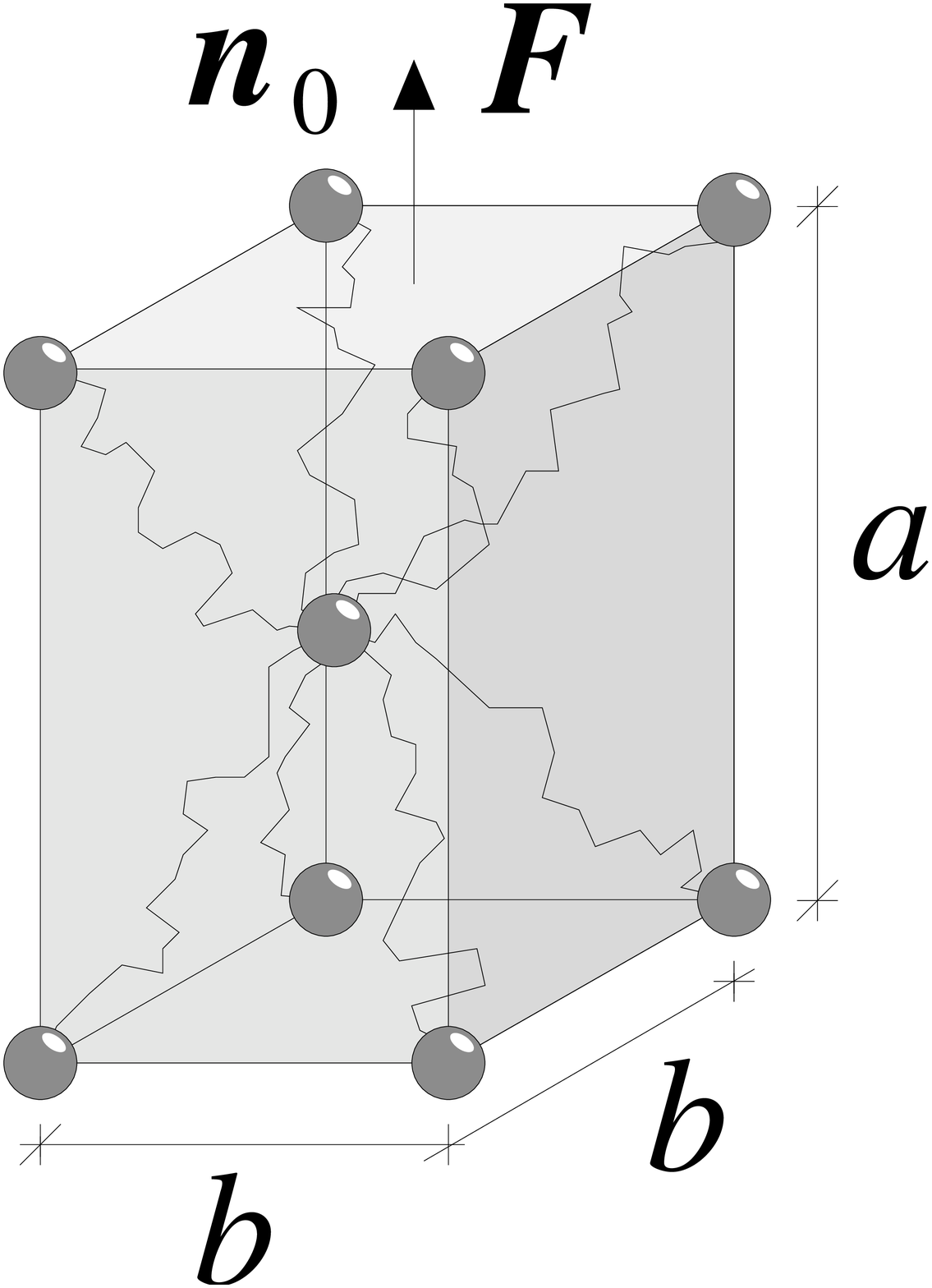}} \put(
69.5,-2.0){\includegraphics[width= 70.mm,angle=0]{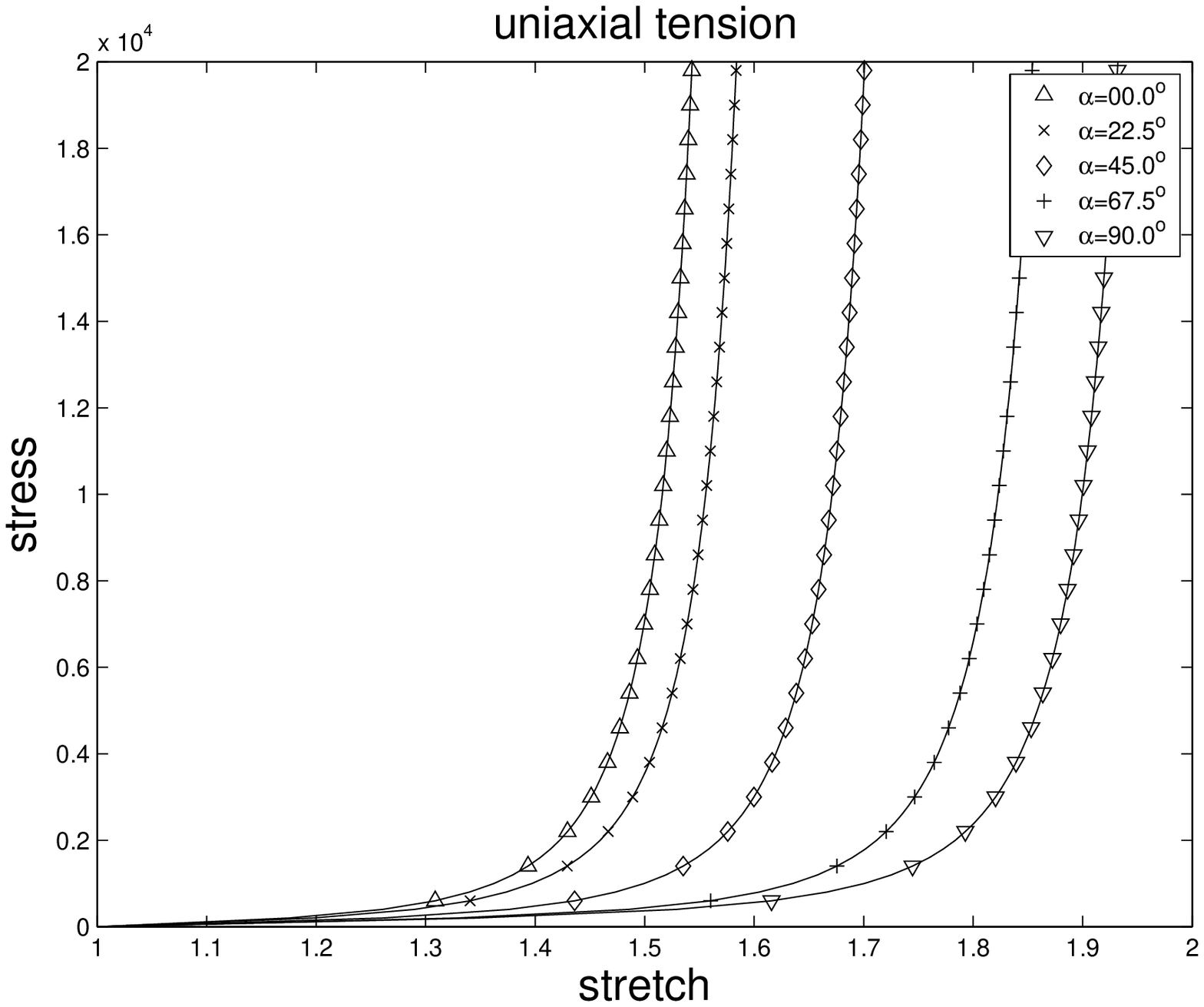}} \put(
77.4,25.0){\includegraphics[width= 17.mm,angle=0]{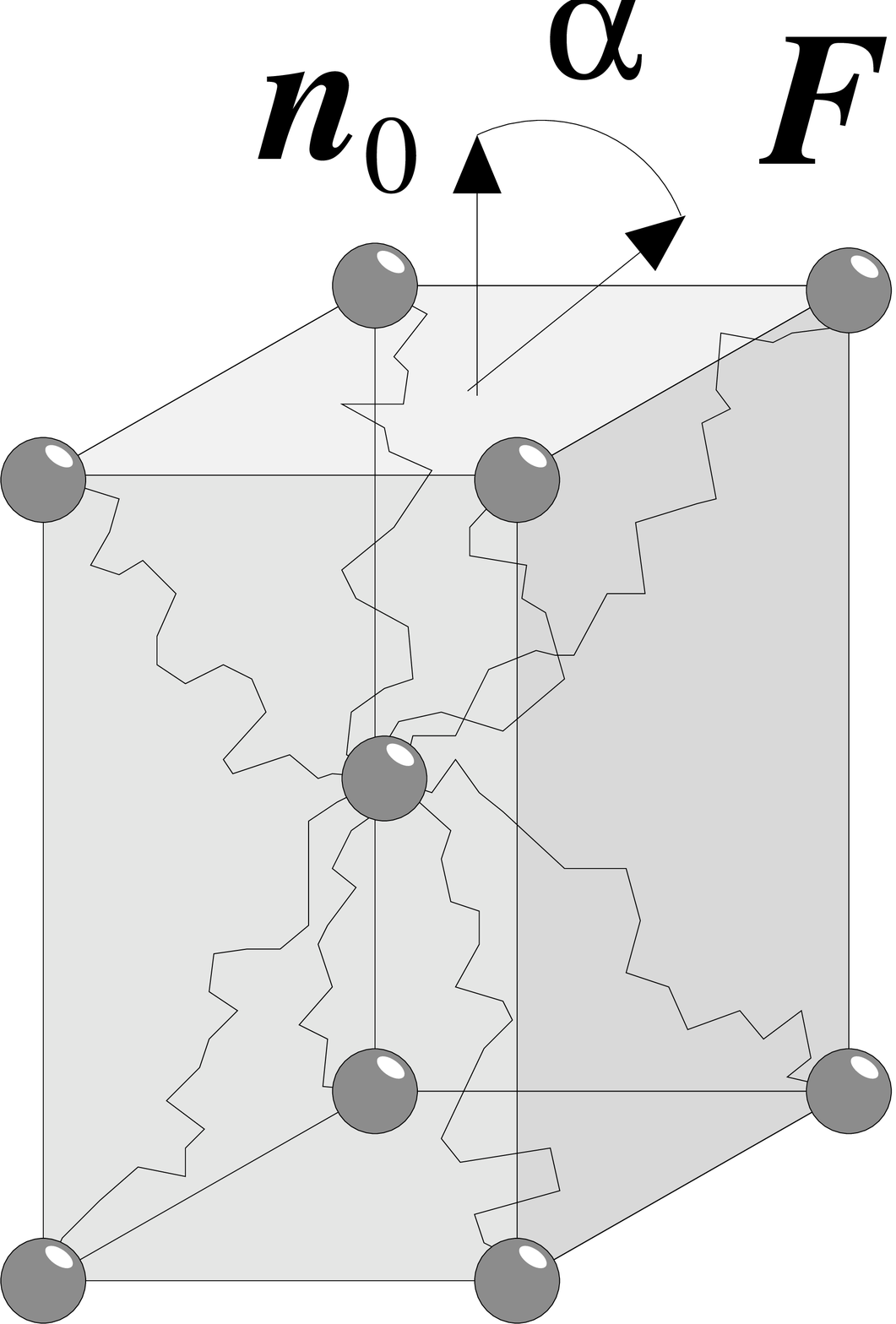}}
\end{picture}
\caption{Influence of anisotropy -- variation of cell dimensions and fiber load
angle}
\label{anisotropy}
\end{figure}\\
%%%%%%%%%%%%%%%%%%%%%%%%%%%%%%%%%%%%%%%%%%%%%%%%%%%%%%%%%%%%%%%%%%%%%%%%%
Figure \ref{anisotropy}, left, shows the influence of the cell dimensions
for different unit cell heights $a$ with $b=1.95$ fixed.
Thereby the loading axis $\vec{F}$ is aligned with the fiber direction
$\vec{n}_0$.
As the cell height increases from $a=1.95$, i.e. $a:b=1.0$, right
curve, to $a=2.925$, i.e. $a:b=1.5$, left curve,
the material stiffens considerably. For this particular choice of
parameters, the material with an aspect ratio of $a:b=1.0$, i.e. the isotropic
material, has a locking stretch which is
slightly larger than $\lambda^*=1.98$ while the locking stress of the material
with $a:b=1.5$ is close to $\lambda^*=1.24$.
Recall that at fixed contour length $L$,
fixed persistence length $A$ and fixed cell dimension
$b$ changes in the cell height $a$ imply changes in the initial
end-to-end length $r_0$ which obviously lead to changes in the overall
stress strain response. The anisotropic material response is
thus highly sensitive to changes in the cell dimensions.\\
Figure \ref{anisotropy}, right, illustrates the influence of the orientation of the
fiber direction $\vec{n}_0$ with respect to the loading axis $\vec{F}$.
For a fixed aspect ratio of $a:b=1.25$ and
cell dimensions $a=2.43$ and $b=1.95$,
different fiber load angles have been studied varying from
$\alpha=0^0$, left curve, to $\alpha=90^0$, right curve. As expected, the
material stiffens as the fiber direction rotates towards the loading axis. For
$\alpha=90^0$, the material behaves most compliant with a locking stretch of
about $\lambda^*=1.93$, while the locking stretch of the stiffest response
at $\alpha=0^0$ is close to $\lambda^*=1.55$.
%%%%%%%%%%%%%%%%%%%%%%%%%%%%%%%%%%%%%%%%%%%%%%%%%%%%%%%%%%%%%%%%%%%%%%%%
\subsection{Example: Anisotropic response of rabbit skin}
%%%%%%%%%%%%%%%%%%%%%%%%%%%%%%%%%%%%%%%%%%%%%%%%%%%%%%%%%%%%%%%%%%%%%%%%
Finally, the transversely isotropic eight chain model will be applied to simulate
the behavior of skin. The following simulation is based on an
experiment carried out by
Lanir \& Fung \cite{lanir74},
see also Fung \cite{fung93} for further details and
Bischoff et al. \cite{bischoff02} for corresponding orthotropic
eight chain model simulations.
In the experiment, samples of rabbit skin have
been tested parallel and orthogonal to the head to tail direction. Since the
collagen fibers in skin are basically oriented along the head to tail axis,
the skin samples' responses were much stiffer in the head to tail direction
than orthogonal to \nolinebreak[4]it.
%%%%%%%%%%%%%%%%%%%%%%%%%%%%%%%%%%%%%%%%%%%%%%%%%%%%%%%%%%%%%%%%%%%%%%%%
\begin{figure}[h]
\unitlength1.0mm
%\begin{picture}(139.0,58.0)
\begin{picture}(139.0,55.0)
\put( -1.0,-2.0){\includegraphics[width=
70.mm,angle=0]{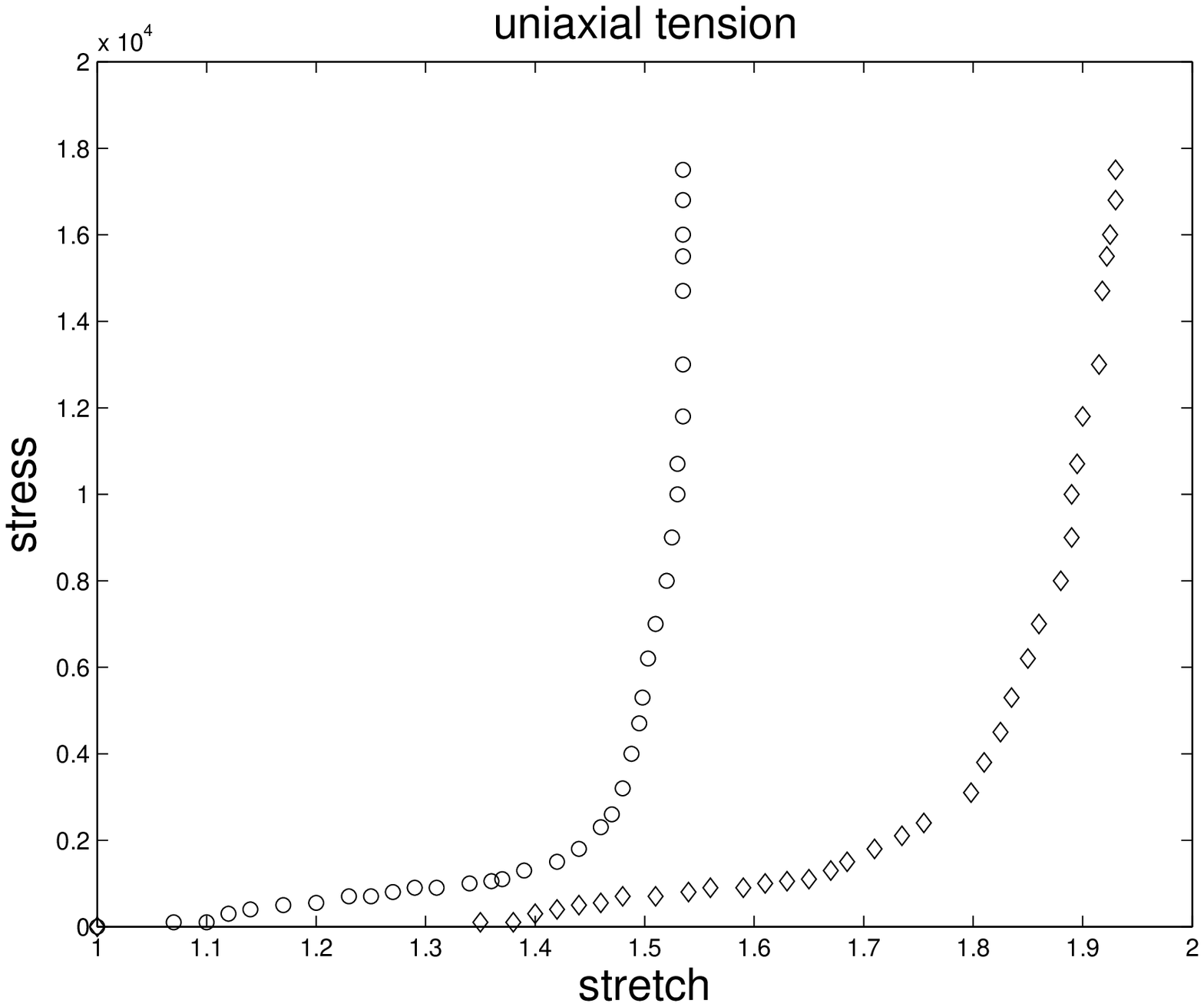}}
%\put(  6.4,21.5){\includegraphics[width= 20.mm,angle=0]{chain_ortho01}}
\put( 69.5,-2.0){\includegraphics[width=
70.mm,angle=0]{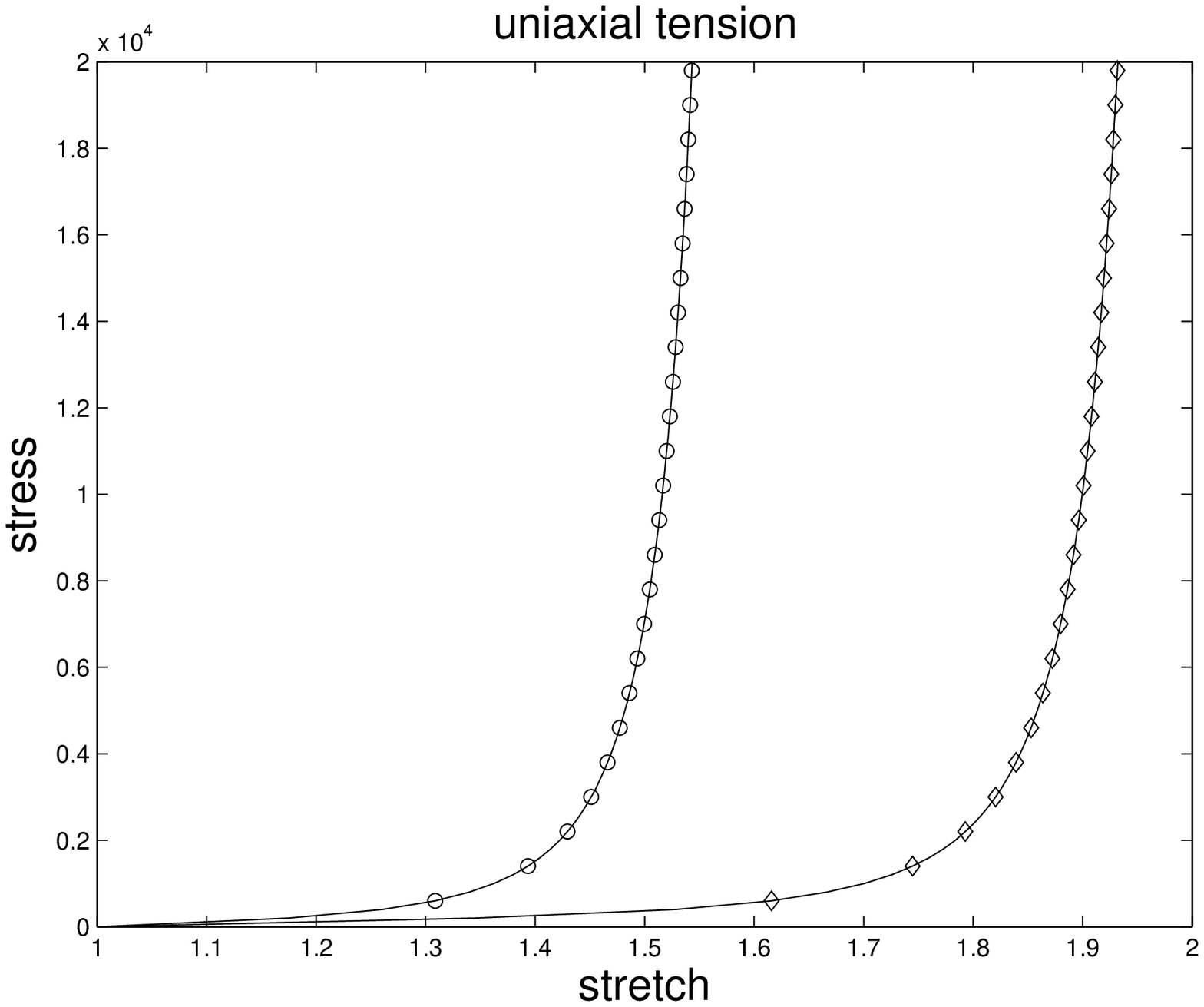}} \put(
77.4,26.0){\includegraphics[width= 18.mm,angle=0]{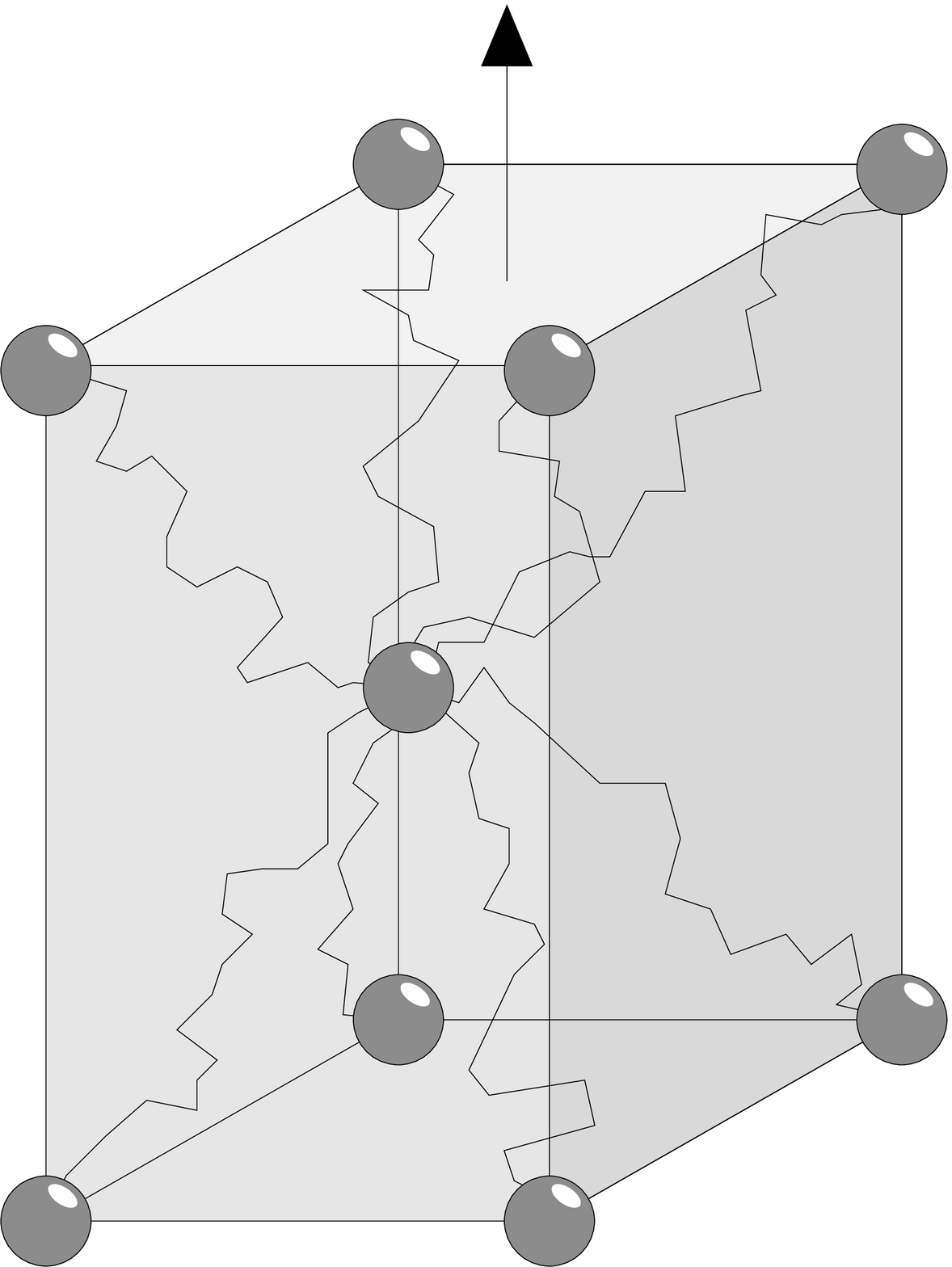}}
\put(26.0,48.0){\footnotesize{$\parallel$ head}}
\put(50.0,48.0){\footnotesize{$\perp$ head}}
\put(29.5,44.0){\footnotesize{tail}}
\put(54.4,44.0){\footnotesize{tail}}
\put(96.0,48.0){\footnotesize{$\parallel$ head}}
\put(120.0,48.0){\footnotesize{$\perp$ head}}
\put(99.5,44.0){\footnotesize{tail}}
\put(124.4,44.0){\footnotesize{tail}}
\end{picture}
\caption{Anisotropy of rabbit skin -- experiment vs. simulation}
\label{rabbit_skin}
\end{figure}\\
%%%%%%%%%%%%%%%%%%%%%%%%%%%%%%%%%%%%%%%%%%%%%%%%%%%%%%%%%%%%%%%%%%%%%%%%%
With the set of parameters introduced in the
previous section,
$L=2.125$, $A=1.82$, $a=2.43$, $b=1.95$, $\gamma^{\scas{chn}}=7 \times 10^{21}$,
$\gamma^{\scas{blk}}=100$ and $\beta=4.5$,
the transversely isotropic wormlike chain model nicely captures the experimental
results by Lanir \& Fung \cite{lanir74} which are given in figure
\ref{rabbit_skin}, left, whereby the data points have been reproduced from the
original paper. Not only the different locking stretches of
about $\lambda^*=1.55$ and $\lambda^*=1.93$ are captured nicely by the
simulation of figure \ref{rabbit_skin}, right, also
the overall shape of the curves corresponds to the experimental findings.
\\[2.ex]
%%%%%%%%%%%%%%%%%%%%%%%%%%%%%%%%%%%%%%%%%%%%%%%%%%%%%%%%%%%%%%%%%%%%%%%%
\begin{remark}[Freely jointed chain network model] \quad
Recall that the use of the freely jointed chain of section
\ref{micro_fjc} within the
eight chain network model typically
overpredicts the locking behavior. Lacking the freedom of the
second characteristic parameter, the freely jointed chain model experiences
difficulties in simultaneously
adjusting the appropriate locking stretch and the shape of the stress
stretch curve.
The two-parameter wormlike chain model, however, nicely captures
both characteristics, not only in the single chain case but also when
embedded in the representative chain network.
\end{remark}
%%%%%%%%%%%%%%%%%%%%%%%%%%%%%%%%%%%%%%%%%%%%%%%%%%%%%%%%%%%%%%%%%%%%%%%%

%%%%%%%%%%%%%%%%%%%%%%%%%%%%%%%%%%%%%%%%%%%%%%%%%%%%%%%%%%%%%%%%%%%%%%%%
\section{Remodeling of the transversely isotropic tissue}\label{remod}
%%%%%%%%%%%%%%%%%%%%%%%%%%%%%%%%%%%%%%%%%%%%%%%%%%%%%%%%%%%%%%%%%%%%%%%%
Soft biological tissues such as muscles, tendons or ligaments show a pronounced
orientation of collagen fibers along one or two particular directions.
Nevertheless, this phenomenon
which is typically not observable in neonatal tissue only develops upon
mechanical loading and is often referred to as functional adaptation.
In the present section, we suggest a theory that allows for a
continuous redistribution of the material's principal directions. For the
transversely isotropic chain network model, we thus allow the
characteristic cell axis to rotate according to a particular
mechanical stimulus, in our case, the maximum principal strain.
%%%%%%%%%%%%%%%%%%%%%%%%%%%%%%%%%%%%%%%%%%%%%%%%%%%%%%%%%%%%%%%%%%%%%%%%
\subsection{Continuum model of remodeling}
\begin{figure}[t]
\unitlength1.0mm
\begin{picture}(139.0,60.0)
\put( -1.0, 0.0){\includegraphics[width=139mm,angle=0]{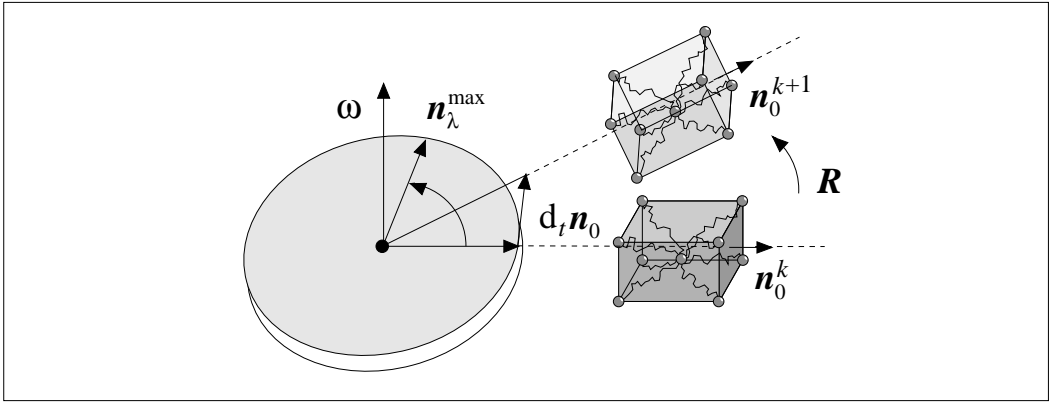}}
\end{picture}
\caption{Reorientation with respect to principal strains}
\label{pic_omega}
\end{figure}

%%%%%%%%%%%%%%%%%%%%%%%%%%%%%%%%%%%%%%%%%%%%%%%%%%%%%%%%%%%%%%%%%%%%%%%%
%$  \sca{D}_t \vec{n}_0
% =[\vec{n}_0 - \vec{n}_{\scas{eq}}  ]/{\tau_{\omega}}$
%with $\vec{n}_{\scas{eq}}$ denoting the biological equilibrium state,
The fundamental assumption of the original isotropic eight-chain model by
Arruda \& Boyce \cite{arruda93,boyce96,boyce00} is that once a particular
deformation is applied, the unit cell is assumed to rotate {\it instantaneously}
towards the principal axes.
The basic idea of the present model is that the unit cell axis
$\vec{n}_0$ is allowed to {\it gradually}
align with eigenvector $\vec{n}_{\lambda}^{\scas{max}}$ of Cauchy Green tensor
$\ten{C}=\ten{F}^{\scas{t}}\cdot \ten{g} \cdot\ten{F}$
which is associated with maximum eigenvalue
${\lambda}^{\scas{max}}=\max \, ( \lambda_i )$.
%whereby $\lambda_3 > \lambda_2 \ge \lambda_1$.
Here, $\vec{n}_{\lambda}^{\scas{max}}$ follows straightforwardly from the
spectral decomposition of $\ten{C}$.
\beq
\ten{C} = \lambda_i \; \vec{n}_{\lambda}^i \otimes \vec{n}_{\lambda}^i
\qquad i=1,2,3
\eeq
In some biologically relevant cases, we might encounter multiple maximum
eigenvalues $\lambda_i$. For the sake of clarity, for the time being,
we assume that no alignment takes place for multiple maximum
eigenvalues.
Following the approach suggested recently by Menzel \cite{menzel04},
we introduce the rotation vector \nolinebreak[4]$\vec{\omega}$
\beq
   \vec{\omega}
= \frac{1}{\tau_{\omega}} \, \vec{n}_0 \times \vec{n}_{\lambda}^{\scas{max}}
\label{omega}
\eeq
as the scaled
vector product of the unit cell orientation $\vec{n}_0$ and the direction of the
maximum principal strain $\vec{n}_{\lambda}^{\scas{max}}$ as sketched in figure \ref{pic_omega}.
%%%%%%%%%%%%%%%%%%%%%%%%%%%%%%%%%%%%%%%%%%%%%%%%%%%%%%%%%%%%%%%%%%%%%%%%
%%%%%%%%%%%%%%%%%%%%%%%%%%%%%%%%%%%%%%%%%%%%%%%%%%%%%%%%%%%%%%%%%%%%%%%
The decomposition of the
rotation vector $\vec{\omega}$
into the
unit normal $\vec{n}^{\omega}$ with
$\vec{n}^{\omega}\cdot\vec{n}^{\omega}=1$
and the magnitude $\omega$ as
\beq
   \vec{\omega}
= \omega \, \vec{n}^{\omega}
  \qquad \;
  \vec{n}^{\omega} =
  \frac{\vec{n}_0 \times \vec{n}_{\lambda}^{\scas{max}}}
       {||\vec{n}_0 \times \vec{n}_{\lambda}^{\scas{max}} ||}
  \qquad \;
  \omega=\frac{||\vec{n}_0 \times \vec{n}_{\lambda}^{\scas{max}} ||}{\tau_{\omega}}
\eeq
illustrates, that the magnitude of rotation is
obviously governed by both
the time relaxation parameter $\tau_{\omega}$ and
the angle between $\vec{n}_0$ and $\vec{n}_{\lambda}^{\scas{max}}$. With the above considerations
at hand, the
evolution of the unit cell axis $\vec{n}_0$ can be expressed
in the following abstract form
\beq
  \sca{d}_t \, \vec{n}_0
= \vec{\omega} \times \vec{n}_0
= -[ \, \ten{e} \cdot \vec{\omega} \, ] \cdot \vec{n}_0
%  \qquad
%  \vec{\omega}
%= \omega \, \vec{n}_{\omega}
%= \frac{1}{\tau_{\omega}} \, \vec{n}_0 \times \vec{n}_{\lambda}^{\scas{max}}
\label{evolution01}
\eeq
whereby $\ten{e}$ denotes the third order permutation symbol.
With the help of equation (\ref{omega})
the above equation can be reformulated in the following, maybe more
illustrative format.
\beq
  \sca{d}_t \, \vec{n}_0
= \frac{1}{\tau_{\omega}} \,
  [\, \vec{n}_{\lambda}^{\scas{max}} - [\, \vec{n}_{\lambda}^{\scas{max}} \cdot \vec{n}_0 \,] \, \vec{n}_0 \,]
\label{evolution02}
\eeq
Note that for the particular evolution equation (\ref{evolution01}),
the orthogonality condition
$\sca{d}_t \vec{n}_0 \cdot \vec{n}_0 = 0$
is valid throughout the remodeling history.
%%%%%%%%%%%%%%%%%%%%%%%%%%%%%%%%%%%%%%%%%%%%%%%%%%%%%%%%%%%%%%%%%%%%%%%%
\subsection{Discrete model of remodeling}
%%%%%%%%%%%%%%%%%%%%%%%%%%%%%%%%%%%%%%%%%%%%%%%%%%%%%%%%%%%%%%%%%%%%%%%%
In the present section, we suggest a discrete numerical solution strategy to
solve the transient equation (\ref{evolution02}) governing the remodeling process.
For its temporal discretization, consider a partition of the time interval of
interest
${\mathcal{T}}$ into a finite number of subintervals
${\mathcal{T}}=\bigcup_{k=0}^{\scas{n_{step}}-1} [t^{k}, t^{k+1}]$
and focus on the typical subinterval
$[t^{k}, t^{k+1}]$
with $\Delta t = t^{k+1} - t^{k}$ denoting the corresponding time increment.
In doing so, we
assume that the unit cell orientation $\vec{n}_0$ is known at $t^{k}$.
The suggested
constitutive integrator of the evolution equation (\ref{evolution02})
is based on an exponential integration scheme which allows for a closed form
representation of Euler-Rodrigues type.
%%%%%%%%%%%%%%%%%%%%%%%%%%%%%%%%%%%%%%%%%%%%%%%%%%%%%%%%%%%%%%%%%%%%%%%%
%\begin{figure}[h]
%\begin{center}
%\resizebox{1.0\textwidth}{!}{
%\unitlength1.0mm
%\begin{picture}(70.0,17.0)
%\put( -1.0, 0.0){\includegraphics[width=35mm,angle=0]{rot_numer}}
%\end{picture}
%}
%\end{center}
%\caption{Remodeling in terms of rotation tensor}
%\label{eight_ortho}
%\end{figure}
%%%%%%%%%%%%%%%%%%%%%%%%%%%%%%%%%%%%%%%%%%%%%%%%%%%%%%%%%%%%%%%%%%%%%%%
\beq
  \vec{n}_{0}^{k+1}
= \exp (- \Delta t \, \ten{e} \cdot \vec{\omega}) \cdot \vec{n}_{0}^{k}
= \ten{R} \cdot \vec{n}_{0}^{k}
\label{update01}
\eeq
In the above equation, we have introduced the proper orthogonal tensor
$\ten{R} (\vec{\omega})=\ten{R} (\omega, \vec{n}^{\omega})$ which is
characterized through the following closed form expression.
\beq
  \ten{R}
= \cos (\Delta t \, \omega) \ten{I}
- \sin (\Delta t \, \omega) \,
   \ten{e} \cdot \vec{n}^{\omega}
+ [\,1-\cos (\Delta t \, \omega)\,] \,
  \vec{n}^{\omega} \otimes \vec{n}^{\omega}
\label{rotation}
\eeq
The combination of equations
(\ref{update01}) and (\ref{rotation})
straightforwardly renders the following update formula for the cell axis
$\vec{n}_{0}^{k+1} $,
\beq
   \vec{n}_{0}^{k+1}
=  \cos (\Delta t \, \vec{\omega}) \vec{n}_{0}^{k}
+  \sin (\Delta t \, \vec{\omega}) \times \vec{n}_{0}^{k}
+ [\,1-\cos (\Delta t \, \vec{\omega})\,] \,
    [\, \vec{n}^{\omega} \cdot \vec{n}_{0}^{k}  \,] \, \vec{n}^{\omega}
\eeq
as in Menzel \cite{menzel04}.
Recall from equation (\ref{omega}),
that the current rotation vector
$\vec{\omega} = \vec{\omega} \, (\vec{n}_{0}^{k+1} )$
is a function of the new direction $\vec{n}_{0}^{k+1}$. Accordingly, the
above equation is actually an implicit update equation in the unit cell
direction $\vec{n}_{0}^{k+1} $ with
dependencies on $\vec{n}_{0}^{k+1} $ on the left and righthand side. In what
follows, however, we shall tacitly assume that
$\vec{n}_{0}^{k+1}  \times \vec{n}^{\omega}
\approx \vec{n}_{0}^{k}  \times \vec{n}^{\omega}$,
such that the rotation vector
$\vec{\omega}$ can be approximated as a function of the known orientation
$\vec{n}_{0}^{k} $. Accordingly, $\vec{\omega}$ can then
be updated explicitly as
\beq
   \vec{\omega}
= \frac{1}{\tau_{\omega}} \, \vec{n}_{0}^{k}  \times \vec{n}_{\lambda}^{\scas{max}}
\eeq
which is a reasonable assumption in the context of the
gradual alignment postulated herein.\\[2.ex]
%%%%%%%%%%%%%%%%%%%%%%%%%%%%%%%%%%%%%%%%%%%%%%%%%%%%%%%%%%%%%%%%%%%%%%%%
\begin{remark}[Classical eight chain model]
\quad
Recall that the unit cell edges of the classical isotropic eight chain
model of  Arruda \& Boyce \cite{arruda93,boyce96,boyce00}
are assumed to align instantaneously with the axes of principal strain.
For isotropic rubber elasticity, this might be a reasonable assumption.
However, it is widely accepted
that biological tissues show a gradual alignment of the material's principal
axes with respect to a mechanical stimulus
due to a cascade of cellular and molecular events involved in the synthesis and
breakdown of collagen that occurs over time.
The suggested remodeling
approach inherently accounts for the successive
reorientation of the anisotropic unit cell with respect to the direction of
maximum principal strain. Accordingly, the final biological equilibrium
state is always characterized through the alignment of the unit cell axes
$\vec{n}_0$ with the maximum principal strain axes $\vec{n}_{\lambda}^{\scas{max}}$.
\end{remark}
%%%%%%%%%%%%%%%%%%%%%%%%%%%%%%%%%%%%%%%%%%%%%%%%%%%%%%%%%%%%%%%%%%%%%%%%
\subsection{Example: Influence of remodeling}
\begin{figure}[h]
\unitlength1.0mm
\begin{picture}(139.0,58.0)
\put( -1.0,-2.0){\includegraphics[width= 70.2mm,angle=0]{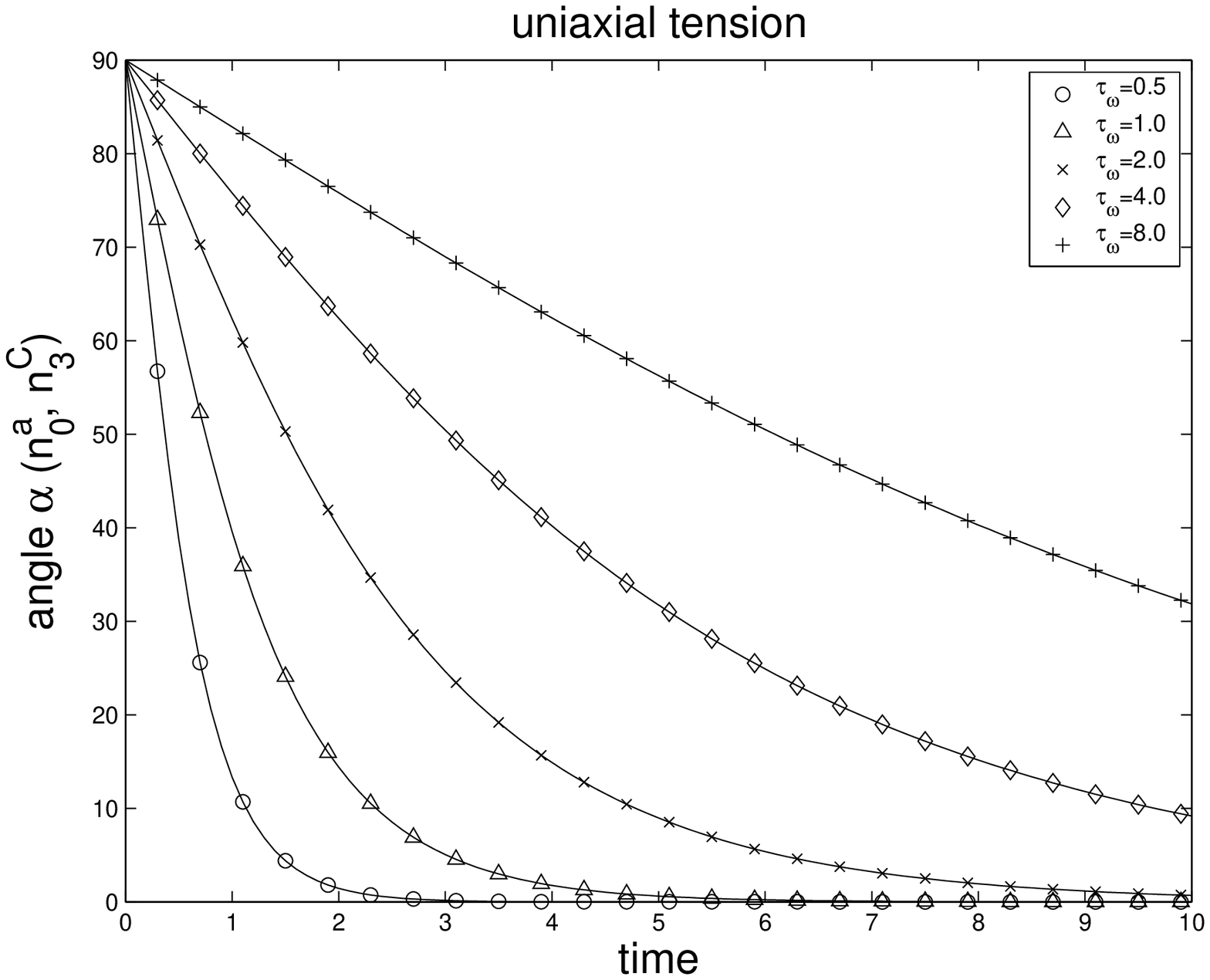}}
\put( 41.0,25.0){\includegraphics[width=
17.mm,angle=0]{chain_ortho03}} \put(
69.5,-2.0){\includegraphics[width= 70.mm,angle=0]{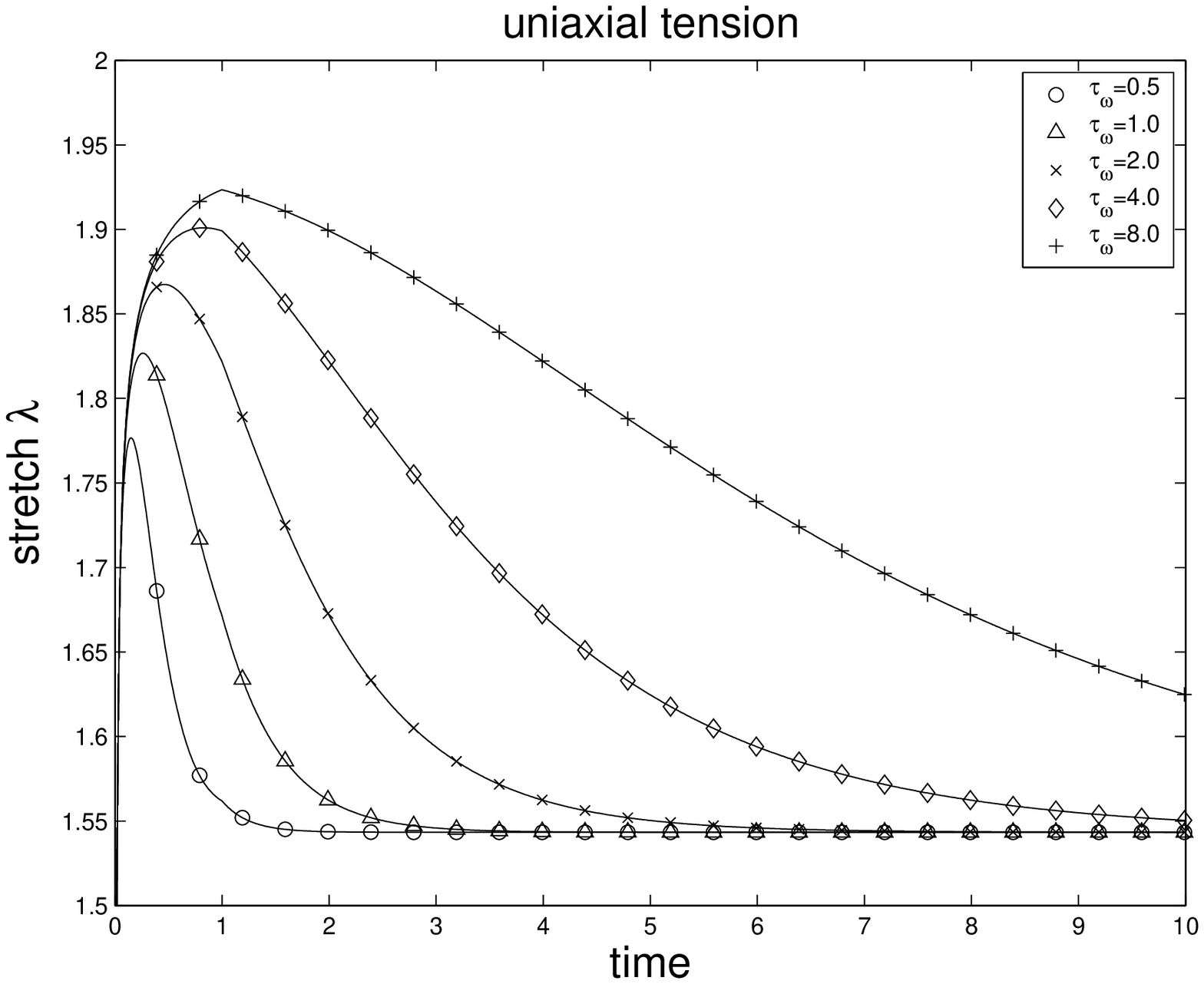}}
\put(109.4,25.0){\includegraphics[width=
17.mm,angle=0]{chain_ortho03}}
\end{picture}
\caption{Remodeling -- reorientation with respect to maximum principal strain
axis}
\label{time}
\end{figure}

%%%%%%%%%%%%%%%%%%%%%%%%%%%%%%%%%%%%%%%%%%%%%%%%%%%%%%%%%%%%%%%%%%%%%%%%
The features of the suggested remodeling approach are demonstrated for the
homogeneous problem under uniaxial tension illustrated in figure
\ref{time}.
The model parameters are adopted from the previous examples of section
\ref{homog}. In addition, the relaxation parameter is chosen to
$\tau = 1.0$ and the time step size is $\Delta t=0.1$.
Similar to the previous examples, the specimen is loaded
up to a final load of $F=20 000$ in the first ten load steps. However, now, the
load is held constant for another 90 time steps to allow for a smooth adaptation
of the cell orientation.
%%%%%%%%%%%%%%%%%%%%%%%%%%%%%%%%%%%%%%%%%%%%%%%%%%%%%%%%%%%%%%%%%%%%%%%%
\\
%%%%%%%%%%%%%%%%%%%%%%%%%%%%%%%%%%%%%%%%%%%%%%%%%%%%%%%%%%%%%%%%%%%%%%%%%
Starting at a fiber load angle of $\alpha=90^0$, the cell axis $\vec{n}_0$
gradually rotates towards the loading axis, i.e. $\alpha=0^0$, as time evolves,
as illustrated in figure \ref{time}, left. The curves
nicely demonstrate the magnitude of adaptation
$\omega=||\vec{n}_0 \times \vec{n}_{\lambda}^{\scas{max}} ||/\tau_{\omega}$
which is obviously large for a large mismatch of axes and which
decreases as the fiber load angle tends to zero. The additional
influence of the relaxation parameter $\tau_{\omega}$ manifests itself in a fast
adaptation for small values of $\tau_{\omega}$, e.g. for
$\tau_{\omega}=0.5$, left curve. For larger values of $\tau_{\omega}$, e.g. for
$\tau_{\omega}=8.0$, right curve, the adaptation time increases. The suggested
model is thus able to capture a gradual reorientation of the axis of transverse
isotropy with respect to the principal strain axis.\\
The local strengthening upon reorientation is nicely documented by figure
\ref{time}, right. It shows the evolution of the stretch as a function of time.
In the loading phase, the stretch increases up to about $\lambda^*=1.93$. This
value corresponds to the locking stretch of the $\alpha=90^0$ orientation in
figure \ref{anisotropy}, right. Then, upon remodeling, the specimen
contracts as the strong material axis rotates towards the loading axis. The
plateau indicates the state of biological equilibrium. The alignment of the cell
axis with the axis of loading reduces the overall stretch to $\lambda^*=1.55$
which agrees nicely with the $\alpha=0^0$ results of figure \ref{anisotropy}, right.
%%%%%%%%%%%%%%%%%%%%%%%%%%%%%%%%%%%%%%%%%%%%%%%%%%%%%%%%%%%%%%%%%%%%%%%%
\subsection{Example: Fiber reorientation in tendons}
%%%%%%%%%%%%%%%%%%%%%%%%%%%%%%%%%%%%%%%%%%%%%%%%%%%%%%%%%%%%%%%%%%%%%%%%
Let us now turn to the simulation of a real biomechanical boundary value
problem motivated by an experiment of
engineered functional tendon constructs as documented by
Calve et al. \cite{calve04}.
Experimental observations confirm, that in the absence of loading,
the in vitro grown tendon constructs
showed the typical characteristics of embryonic tendon
demonstrated by the absence of a collageneous scaffold.
It is only upon mechanical loading that collagen fibrils
form within the tendon and orient themselves along the loading
direction. The long term goal of the present project
is thus to experimentally analyze and computationally predict
mechanically stimulated remodeling in the form of fiber reorientation.
In the future, other microstructural changes such as the increase in cross linkage or
the post-depositional fusion of fibrils will be considered as well.
These might be incorporated straightforwardly in the present framework
through changes in the contour length or in the persistence length for
which appropriate evolution equations have to be defined.
%%%%%%%%%%%%%%%%%%%%%%%%%%%%%%%%%%%%%%%%%%%%%%%%%%%%%%%%%%%%%%%%%%%%%%%%
\begin{figure}[h]
\unitlength1.0mm
\begin{picture}(139.0,58.0)
\put( -1.0,-2.0){\includegraphics[width=
72.0mm,angle=0]{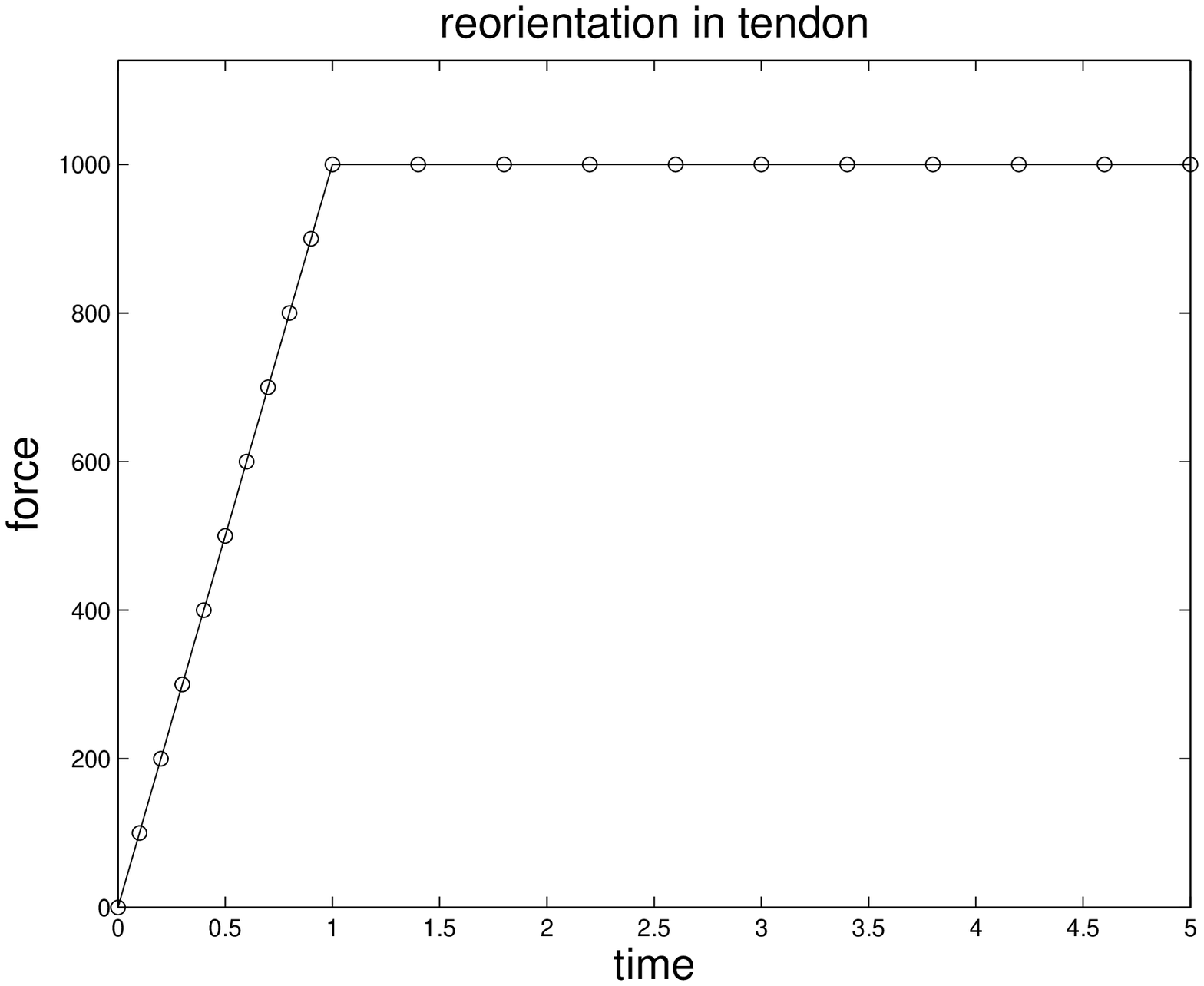}}
%\put( 58.0,8.0){\includegraphics[width= 25.mm,angle=90]{tendon_mod}}
\put(55.5,5.4){\includegraphics[height=30.mm,angle=0]{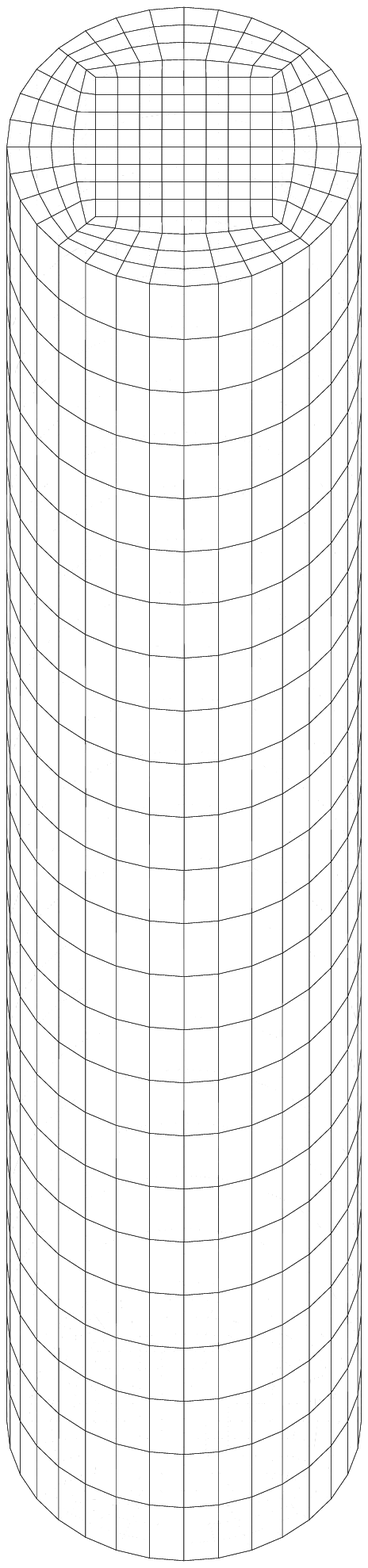}}
%\put( 36.0,16.0){\includegraphics[height=30.mm,angle=-90]{tendon_diag}}
\put( 70.0,-2.0){\includegraphics[width= 70.mm,angle=0]{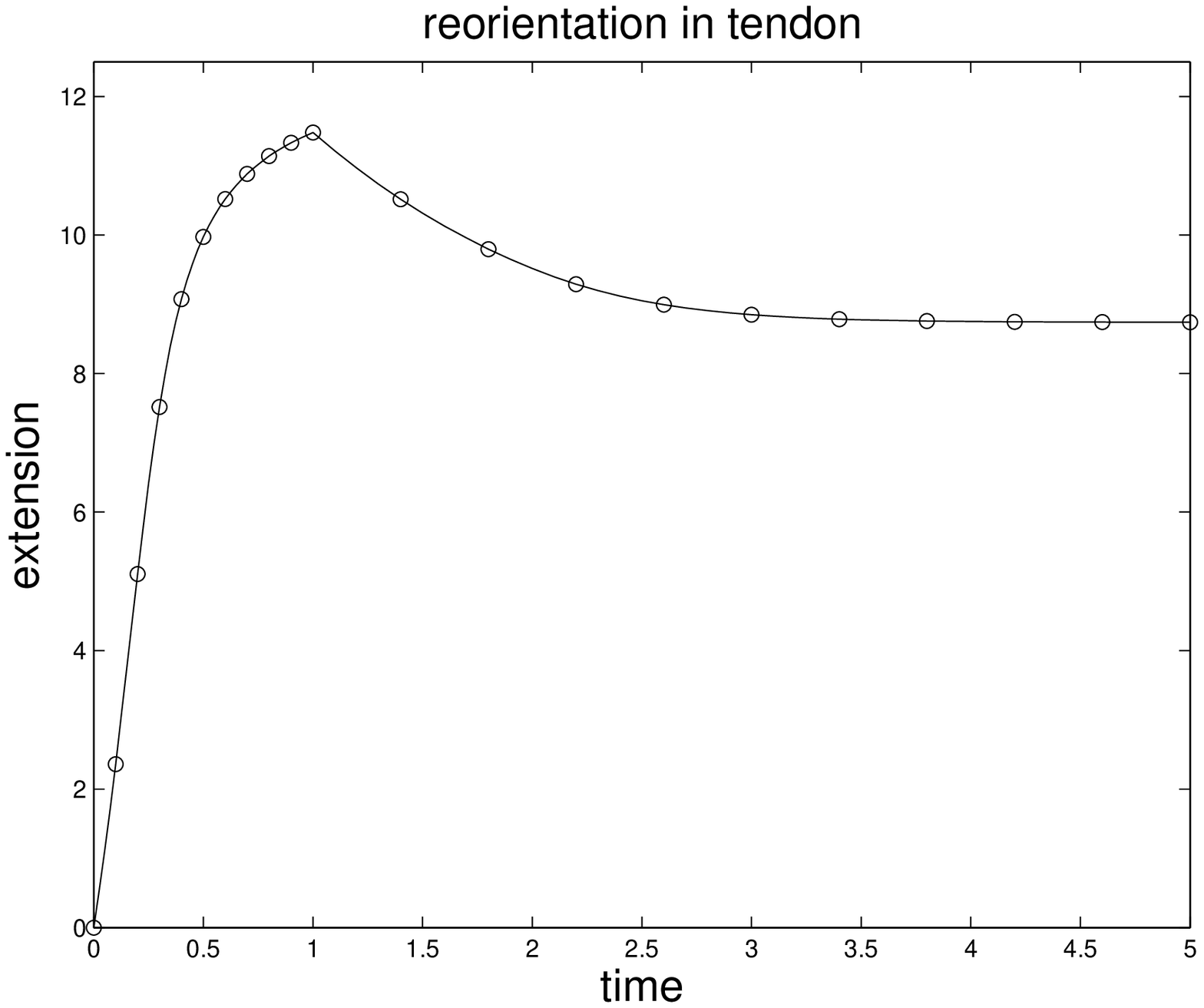}}
%\put(126.0,8.0){\includegraphics[width= 25.mm,angle=90]{tendon_mod}}
\put(125.0,5.4){\includegraphics[height=30.mm,angle=0]{tendon_diag}}
\end{picture}
\caption{Remodeling of soft biological tissue -- prescribed loading and extension}
\label{tendon_curves}
\end{figure}\\
%%%%%%%%%%%%%%%%%%%%%%%%%%%%%%%%%%%%%%%%%%%%%%%%%%%%%%%%%%%%%%%%%%%%%%%%%
As a preliminary study, we carry out a finite element simulation of the
remod\nolinebreak[4]eling process in a cylindrical model tendon. The tendon material is
described with the transversely isotropic wormlike chain model
whereby an initially random fiber orientation is assumed to represent
the neonatal state. The current fiber orientation $\vec{n}_0$ is introduced
as an internal variable which is stored locally
on the integration point level.
The tendon structure has an initial cross section of unit area and a length of
twelve units, respectively. It is discretized with
2304 eight-noded brick elements introducing about
8000 degrees of freedom.
The two wormlike chain parameters, i.e. the
the contour length and the persistence length,
take values of $L=2.50$ and $A=1.82$.
Note that all lengths in the model are
normalized by the segment length $l$. The
bulk parameters take the values of $\gamma^{\scas{blk}}=100$ and $\beta=4.5$,
the chain density is chosen to $\gamma^{\scas{chn}}=7 \times 10^{21}$,
the absolute temperature is $\theta=310$. The dimensions of the transversely
isotropic unit cell are taken as $a=2.43$ and $b=1.85$, again normalized with
respect to the segment length $l$.
The relaxation time for the remodeling procedure is chosen as $\tau$=1.0. \\
In the first 40 time steps of $\Delta t =0.0025$,
the model tendon is gradually stretched to about $200$ \% of its initial length.
The final load of $F=1000$ is then held constant
for another 40 time steps of $\Delta t =0.01$
to allow for fiber reorientation, see figure \ref{tendon_curves}, left.
Figure \ref{tendon_curves}, right, shows the temporal evolution of the elongation
of the tendon. As the tendon is loaded, i.e. for $0.0 \le t \le 1.0$, the elongation
increases smoothly up to almost $12$ units. During the reorientation period,
i.e. for $1.0 \le t \le 5.0$,
the tendon obviously stiffens considerably due to fiber reorientation.
Accordingly, the elongation reduces to about $8.5$ units at the
final biological equilibrium state, see figure \ref{tendon_curves}, right.
%%%%%%%%%%%%%%%%%%%%%%%%%%%%%%%%%%%%%%%%%%%%%%%%%%%%%%%%%%%%%%%%%%%%%%%%
\begin{figure}
\unitlength1.0mm
\begin{picture}(139.0,210.0)
\put(15.6,
20.0){\includegraphics[height=24.5mm,width=107.mm,angle=180]{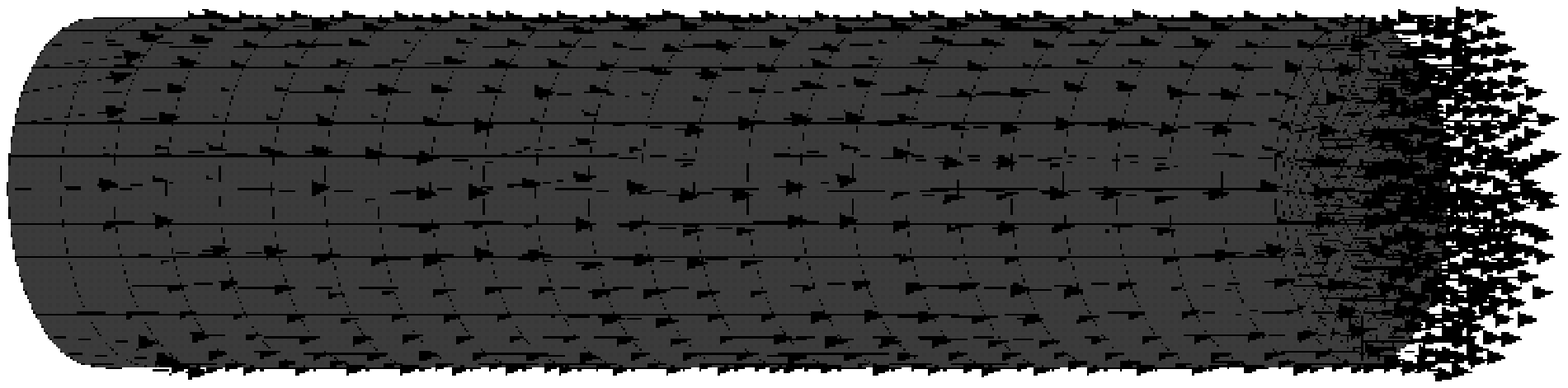}}
\put(15.6,
51.0){\includegraphics[height=24.5mm,width=112.mm,angle=180]{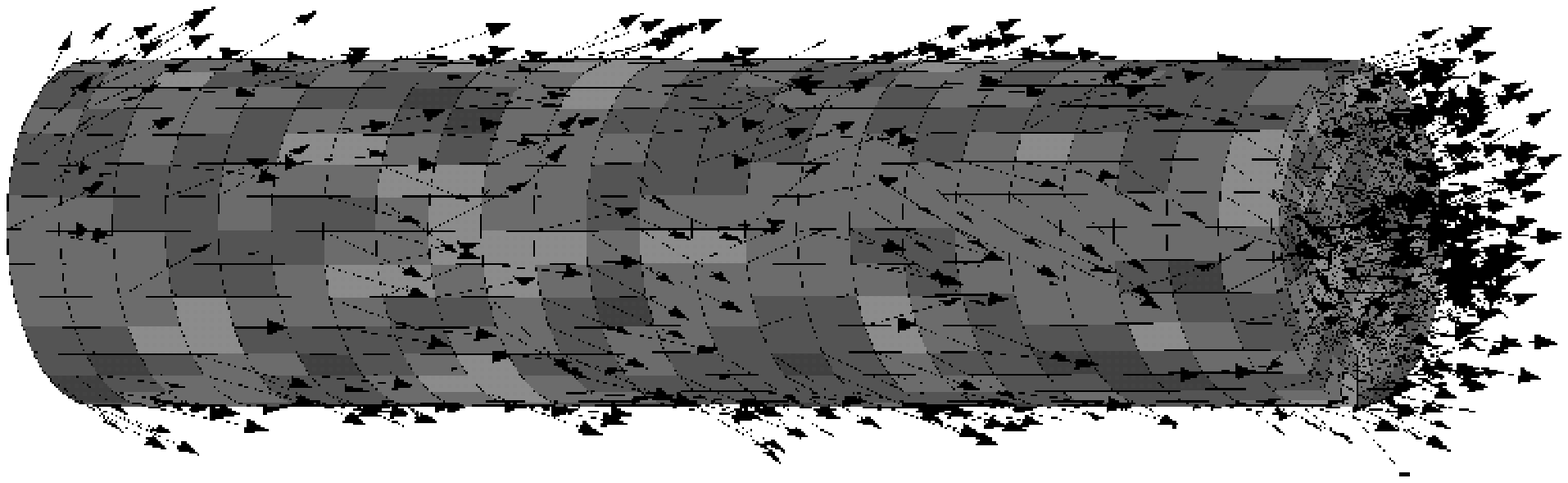}}
\put(15.0,
93.0){\includegraphics[height=42.mm,width=124.mm,angle=180]{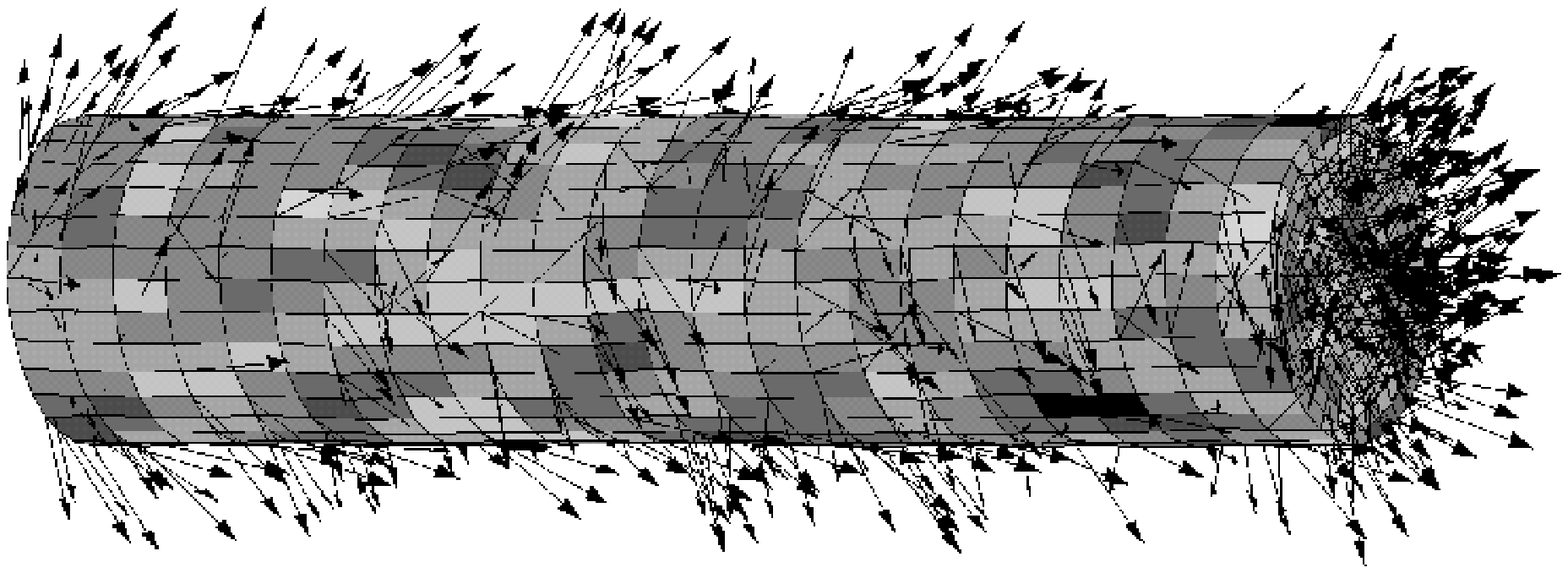}}
\put(15.0,128.0){\includegraphics[height=42.mm,width=124.mm,angle=180]{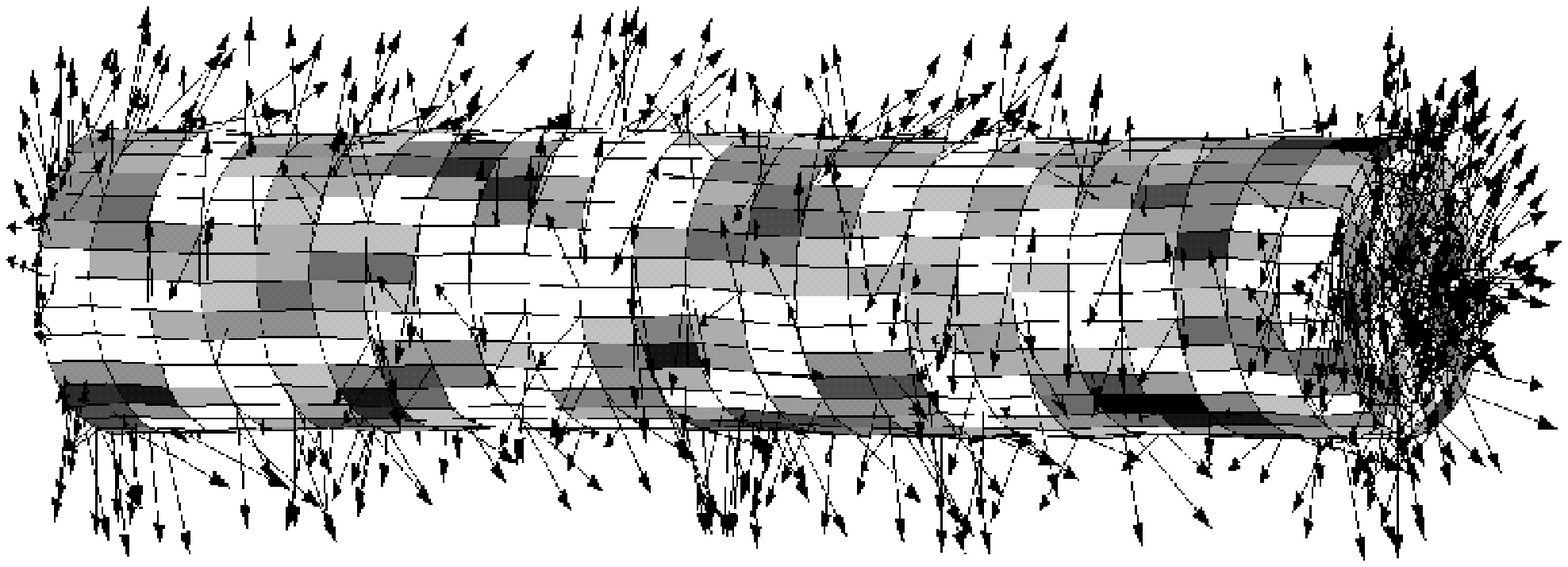}}
\put(15.0,163.0){\includegraphics[height=42.mm,width=124.mm,angle=180]{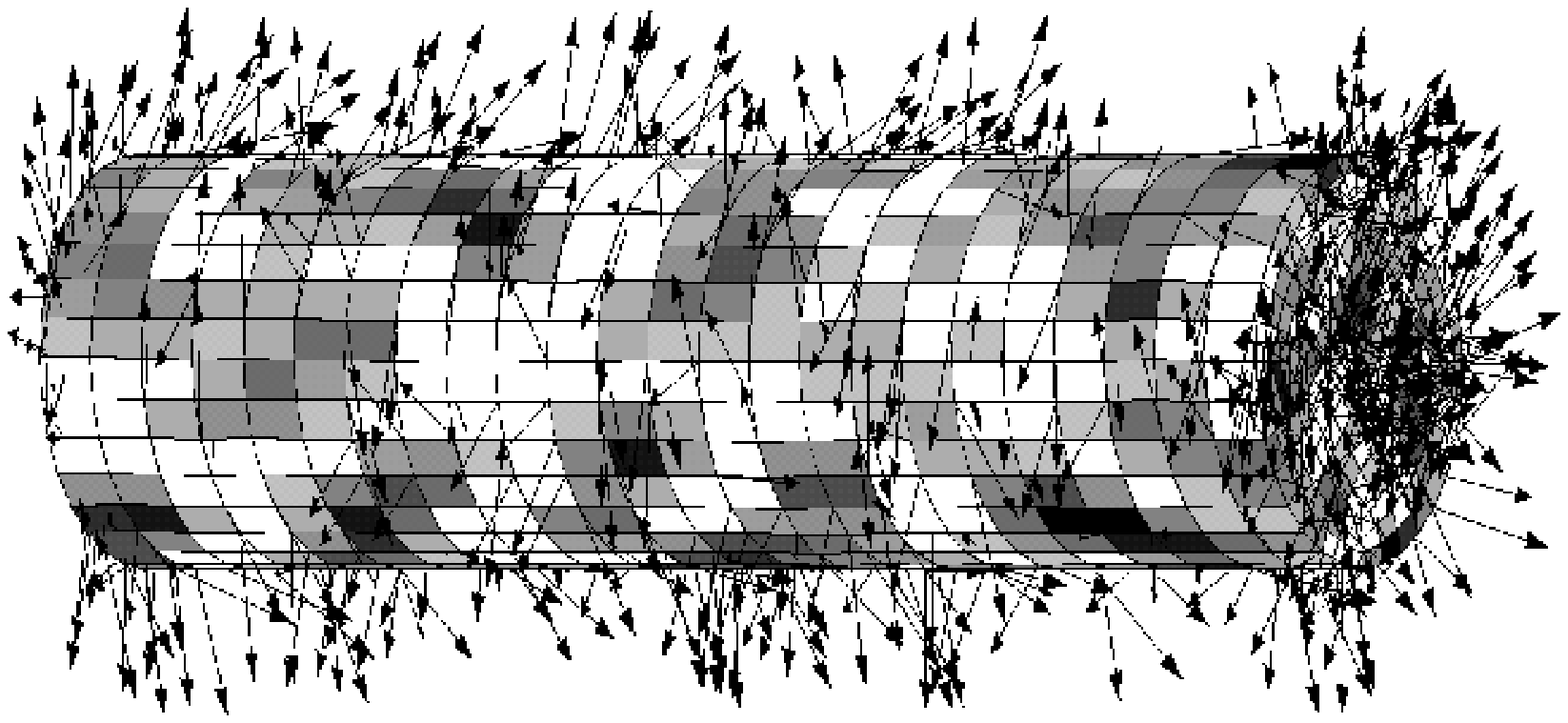}}
\put(15.0,201.0){\includegraphics[height=42.mm,width=124.mm,angle=180]{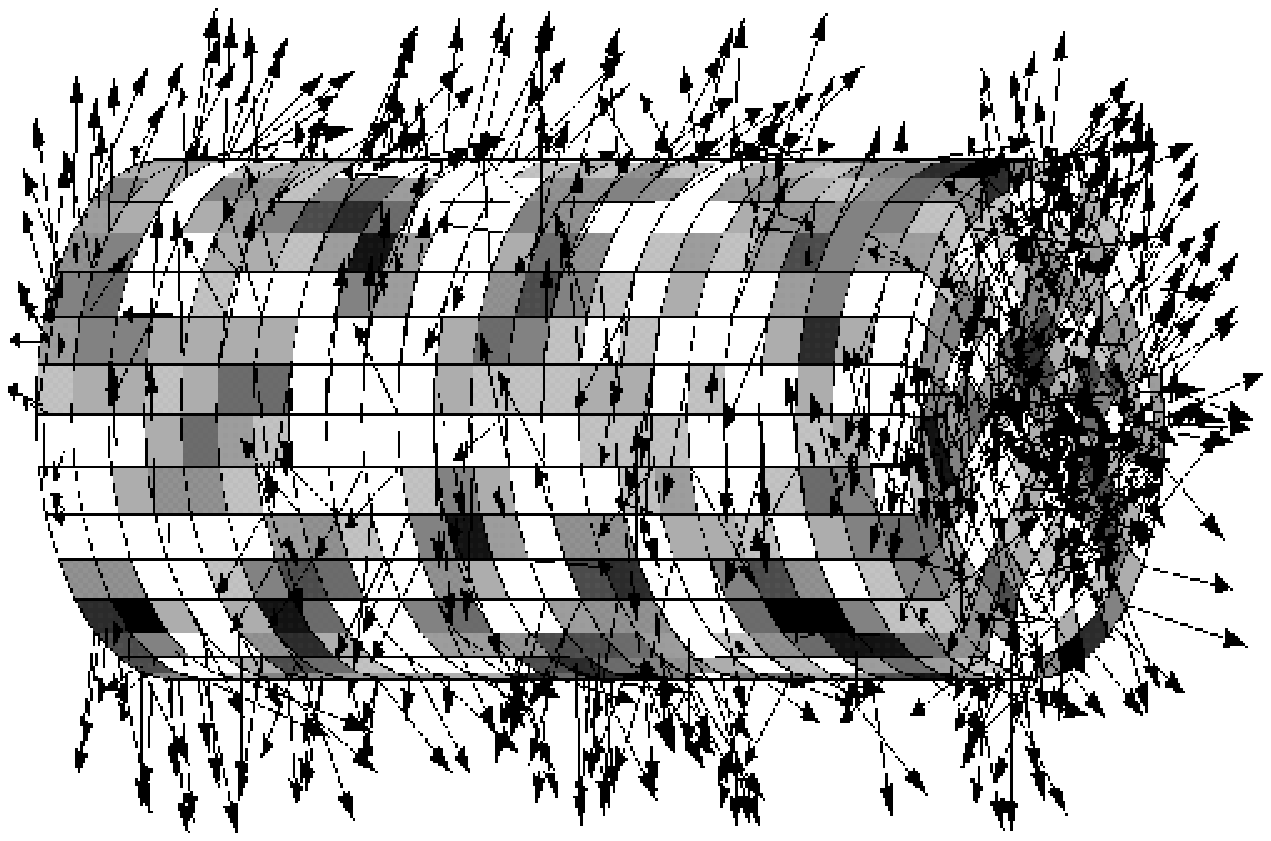}}
\put(0.6,  7.1){t=5.00} \put(0.6, 38.5){t=3.00} \put(0.0,
71.5){t=2.00} \put(0.0,106.0){t=1.00} \put(0.0,141.4){t=0.25}
\put(0.0,179.0){t=0.00} \put(136.0,49.0){$\alpha$}
\put(124.4,-8.4){\includegraphics[height=62.mm,width=20.mm,angle=0]{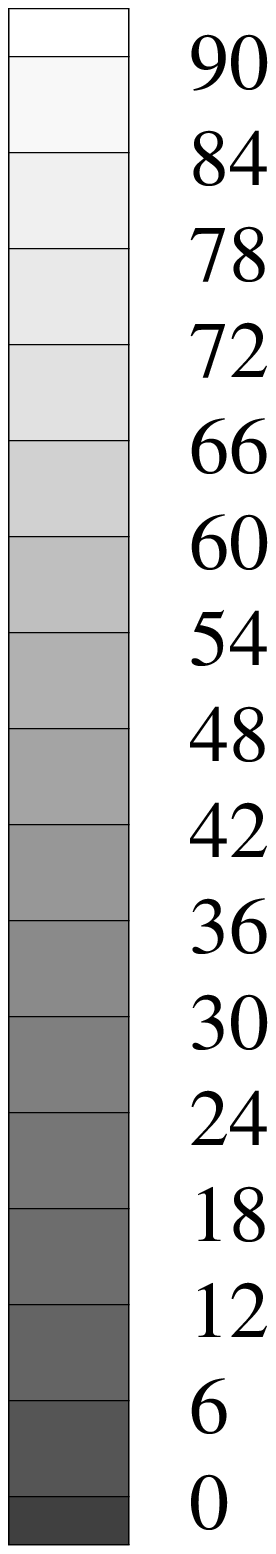}}
%\put(50.6, 16.0){\includegraphics[height=23.mm,width= 95.mm,angle=180]{tendon80a}}
%\put(40.6, 44.0){\includegraphics[height=23.mm,width=100.mm,angle=180]{tendon60a}}
%\put(-134.0,10.0){\includegraphics[width=160.mm,angle=  0]{angle}}
%\put( 86.0,244.0){\includegraphics[width=160.mm,angle=  0]{angle}}
\end{picture}
\caption{Remodeling of soft biologial tissue --
Strain based fiber reorientation of initially randomly oriented
collagen fibers in a cylindrical tendon}
\label{tendon_history}
\end{figure}\\
%%%%%%%%%%%%%%%%%%%%%%%%%%%%%%%%%%%%%%%%%%%%%%%%%%%%%%%%%%%%%%%%%%%%%%%%
Figure \ref{tendon_history} depicts six representative stages of the
remodeling history. The top figure shows the initial state at the
beginning of the loading history. The arrows indicate the the initial
orientation of the unit cells $\vec{n}_0$, or, in the biomechanical sense,
the directions of pronounced collagen fiber orientation,
that have been assigned randomly to each individual integration point. The
contour levels ranging from white to black depict the fiber load angle $\alpha$
varying from $\alpha=90^0$, i.e. the strong cell axis being orthogonal to the
loading axis, to $\alpha=0^0$, i.e. a full
alignment of the cell axis with the axis of loading.\\
Figure \ref{tendon_history} from top to bottom nicely documents the
history of remodeling. While the tendon is loaded uniaxially, a clear
reorientation of the collagen fibers with respect to the loading axis
can be observed. In this sense, the last figure of the series represents a
biological equilibrium state, for which $\sca{d}_t \, \vec{n}_0 = \vec{0}$. No
further remodeling takes place as all fibers are aligned with the maximum
principal strain direction. Accordingly, the fiber load angle indicated through
the underlying contour plots
evolves gradually from a random distribution in the top figure
to a uniform distribution with a fiber load angle of $\alpha=0^0$ in the bottom
figure.\\
While converging towards the final biological equilibrium state
the suggested procedure showed a remarkably stable algorithmic behavior.
Recall that during the loading process,
the tendon has been stretched to twice its original length. The stretch of
the tendon as well as the remarkable reduction of its cross section are clearly
visible in figure \ref{tendon_history}.
The algorithm is thus able to capture both,
kinematic and constitutive non-linearities.
Apparently, the explicit update of the fiber direction does not lead to
computational instabilities as long as the relaxation parameter is sufficiently large
and the time step size is chosen sufficiently small.

%%%%%%%%%%%%%%%%%%%%%%%%%%%%%%%%%%%%%%%%%%%%%%%%%%%%%%%%%%%%%%%%%%%%%%%%
\section{Discussion}\label{concl}
%%%%%%%%%%%%%%%%%%%%%%%%%%%%%%%%%%%%%%%%%%%%%%%%%%%%%%%%%%%%%%%%%%%%%%%%
A new transversely isotropic chain network model has been proposed which is
particularly suited to simulate remodeling of collagen fibers in soft biological
tissue. For the individual chains, a wormlike chain model was adopted. In
contrast to the classical uncorrelated freely jointed chain, the correlated
wormlike chain shows an initial stiffness such that its curvature changes
smoothly in the contour space. The stiffness, or rather the degree of
correlation, is governed by a second parameter besides the contour length, the
so-called persistence length. With the additional freedom of this second
parameter, the wormlike chain model can capture the behavior of the freely
jointed chain and a beam-like behavior as limit cases.\\
The chain network is characterized by a representative eight chain unit cell. In
contrast to the cubic cell of the isotropic eight-chain model,
the present unit cell has one preferred direction which characterizes the axis
of transverse isotropy. As such, the present model can be understood as a
generalization of the classical eight chain model which captures the
original isotropic eight chain model as a special limit case with equal cell
dimensions. In the other limit, i.e. when the in-plane dimensions of the
unit cell tend to zero, the model captures the effects of unidirectional fiber
reinforcement neglecting cross-link effects of the network structure.\\
To incorporate biomechanical remodeling, the characteristic axis of the
transversely isotropic unit cell was allowed to rotate with respect to a
biological stimulus, in our case the maximum principal strain axis. A
theoretical framework of remodeling and its numerical realization based on an
exponential update scheme of Euler-Rodrigues type have been introduced. Thereby,
the characteristic unit cell axis has been introduced as an internal variable on
the integration point level of a finite element realization.\\
A first comparison of the suggested approach with the experimental findings of
in vitro engineered tendon constructs revealed a remarkably good agreement.
Being essentially based on micromechanical considerations, the present model is
governed by a limited number of physically-motivated material parameters. As
such, it is believed to be ideally suited to simulate not only the passive
behavior of soft biological tissues but also their active response to changes in
the mechanical loading environment.

%%%%%%%%%%%%%%%%%%%%%%%%%%%%%%%%%%%%%%%%%%%%%%%%%%%%%%%%%%%%%%%%%%%%%%%%%
\bibliography{litra}
\bibliographystyle{ellen}
%%%%%%%%%%%%%%%%%%%%%%%%%%%%%%%%%%%%%%%%%%%%%%%%%%%%%%%%%%%%%%%%%%%%%%%%%
\end{document}